\documentclass[iop]{emulateapj}
\usepackage{amsmath,amssymb}
\usepackage{xspace}
\usepackage{graphicx}
\usepackage{graphicx}
\usepackage{epsfig}

\def\hst{{\it HST\/}}       

\def\bandu{\textit{U}} 
\def\bandb{\textit{B}}
\def\bandv{\textit{V}}
\def\bandr{\textit{R}}
\def\bandi{\textit{I}}
\def\bandzc{\textit{zc}}
\def\bandzo{\textit{zo}}
\def\bandhk{\textit{HK$^\prime$}}
\def\bandj{\textit{J}}
\def\bandh{\textit{H}}
\def\bandk{\textit{K$_s$}}
\def\aone{\textit{A1}}
\def\atwo{\textit{A2}}
\def\sone{\textit{S1}}
\def\stwo{\textit{S2}}
\def\sthr{\textit{S3}}
\def\sfor{\textit{S4}}

\def\zs{\textit{z$_{\rm spec}$}}
\def\zp{\textit{z$_{\rm phot}$}}
\def\z{\textit{z}}

\def\za{\textit{z$_{\rm alt}$}}
\def\ze{\textit{z$_{\rm peak}$}}
\def\one{\uppercase\expandafter{\romannumeral1}}
\def\two{\uppercase\expandafter{\romannumeral2}}
\def\thr{\uppercase\expandafter{\romannumeral3}}

\def\xray{\hbox{X-ray}}  
\def\hdfn{\hbox{H-HDF-N}}
\def\goodsn{\hbox{GOODS-N}}
\def\goodss{\hbox{GOODS-S}}
\def\cdfn{\hbox{CDF-N}}

\def\lsim{\mathrel{\rlap{\lower4pt\hbox{\hskip1pt$\sim$}}
    \raise1pt\hbox{$<$}}}                
\def\gsim{\mathrel{\rlap{\lower4pt\hbox{\hskip1pt$\sim$}}
    \raise1pt\hbox{$>$}}}                
\def\AA{\buildrel _{\circ} \over {\mathrm{A}}}   
\newcommand{\degree}{\ensuremath{^\circ}}
\def\degree{\ensuremath{^\circ}}
\def\sigm{$\rm \sigma_{NMAD}$}

\usepackage{hyperref} 
\usepackage{xcolor}

\slugcomment{}
\shorttitle{Photo-z in H-HDF-N}
\shortauthors{Yang et~al.}

\begin{document}
\title{Photometric Redshifts in the Hawaii-Hubble Deep Field-North (\hdfn)}
\author{G. Yang\altaffilmark{1},
Y.~Q. Xue\altaffilmark{1},
B. Luo\altaffilmark{2,3},
W.~N. Brandt\altaffilmark{2,3}, 
D.~M. Alexander\altaffilmark{4},
F.~E. Bauer\altaffilmark{5,6,7},
W. Cui\altaffilmark{8},
X. Kong\altaffilmark{1},
B.~D. Lehmer\altaffilmark{9,10},
J.-X. Wang\altaffilmark{1},
X.-B. Wu\altaffilmark{11},
F. Yuan\altaffilmark{12},
Y.-F. Yuan\altaffilmark{1}, and
H.~Y. Zhou\altaffilmark{13,1}
}
\altaffiltext{1}{Key Laboratory for Research in Galaxies and Cosmology, Center for Astrophysics, Department of Astronomy, University of Science and Technology of China, Chinese Academy of Sciences, Hefei, Anhui 230026, China; yg1991@mail.ustc.edu.cn, xuey@ustc.edu.cn}
\altaffiltext{2}{Department of Astronomy and Astrophysics, 525 Davey Lab, The Pennsylvania State University, University Park, PA 16802, USA}
\altaffiltext{3}{Institute for Gravitation and the Cosmos, The Pennsylvania State University, University Park, PA 16802, USA}
\altaffiltext{4}{Department of Physics, Durham University, Durham DH1 3LE, UK}
\altaffiltext{5}{Instituto de Astrof\'{\i}sica, Facultad de F\'{i}sica, Pontificia Universidad Cat\'{o}lica de Chile, 306, Santiago 22, Chile}
\altaffiltext{6}{Millennium Institute of Astrophysics}
\altaffiltext{7}{Space Science Institute, 4750 Walnut Street, Suite 205, Boulder, Colorado 80301}
\altaffiltext{8}{Department of Physics, Purdue University, West Lafayette, IN 47907, USA}
\altaffiltext{9}{The Johns Hopkins University, Homewood Campus, Baltimore, MD 21218, USA}
\altaffiltext{10}{NASA Goddard Space Flight Centre, Code 662, Greenbelt, MD 20771, USA}
\altaffiltext{11}{Department of Astronomy, Peking University, Beijing 100871, China}
\altaffiltext{12}{Key Laboratory for Research in Galaxies and Cosmology, Shanghai Astronomical Observatory, Chinese Academy of Sciences, 80 Nandan Road, Shanghai 200030, China}
\altaffiltext{13}{Polar Research Institute of China, 451 Jinqiao Road, Shanghai, 200136, China}

\begin{abstract}

We derive photometric redshifts (\zp) for sources in the entire ($\sim0.4$ deg$^2$) Hawaii-Hubble Deep Field-North (\hdfn) field with the EAzY code, based on point spread function-matched photometry of 15 broad bands from the ultraviolet (\bandu~band) to mid-infrared (IRAC 4.5 $\mu$m). Our catalog consists of a total of 131,678 sources. We evaluate the \zp~quality by comparing \zp~with spectroscopic redshifts (\zs) when available, and find a value of normalized median absolute deviation \sigm$=$0.029 and an outlier fraction of 5.5\% (outliers are defined as sources having $\rm |\zp - \zs|/(1+\zs) > 0.15$) for non-X-ray sources. More specifically, we obtain \sigm$=0.024$ with 2.7\% outliers for sources brighter than $R=23$~mag, \sigm$=0.035$ with 7.4\% outliers for sources fainter than $R=23$~mag, \sigm$=$0.026 with 3.9\% outliers for sources having $z<1$, and \sigm$=$0.034 with 9.0\% outliers for sources having $z>1$. 
Our \zp\ quality shows an overall improvement over an earlier \zp\ work that focused only on the central \hdfn\ area.
We also classify each object as star or galaxy through template spectral energy distribution fitting and complementary morphological parametrization, resulting in 4959 stars and 126,719 galaxies. Furthermore, we match our catalog with the 2~Ms {\it Chandra} Deep Field-North main \xray~catalog. For the 462 matched non-stellar \xray~sources (281 having \zs), we improve their \zp~quality by adding three additional AGN templates, achieving \sigm$=0.035$ and an outlier fraction of 12.5\%. We make our catalog publicly available presenting both photometry and \zp, and provide guidance on how to make use of our catalog. 
\end{abstract}


\section{Introduction}
Redshifts and consequent results (e.g., luminosity distance, look-back time, and angular distance) are the basis of nearly all observational astronomical studies of extragalactic objects (e.g., luminosity function, mass function, large-scale structures, and galaxy evolution), progress on which would be greatly hampered by the lack of redshift information. The most reliable way to obtain secure redshifts is by taking spectra and identifying emission (or absorption) lines. However, this approach of obtaining spectroscopic redshifts (\zs) is observation time demanding, and proves to be quite challenging especially for very faint objects, e.g., 
the Caltech Faint Galaxy Redshift Survey is 92\% complete down to \bandr=24~mag in the Hubble Deep Field-North and to \bandr=23~mag in the flanking fields (Cohen et al.\ 2000);
however, when targeting slightly fainter objects over a larger area, the Team Keck Treasury Redshift Survey obtained secure spectroscopic redshifts for only 53\% of their targets that are brighter than $\bandr=24.4$~mag in the \goodsn~field (Wirth et al. 2004).

Therefore another approach, obtaining photometric redshifts (\zp) of good quality, is of great value. Determining photometric redshifts with broad-band and/or medium-band imaging observations is able to capture very faint objects in a time-efficient manner, e.g., Dahlen et al.\ (2010) reached a 5$\sigma$ detection limit of ({\it HST}/ACS F850LP) $z=28.1$~mag in their \zp~catalog in the \goodss\ field. Generally speaking, there are two classes of methods for calculating photometric redshifts: empirical and template spectral energy distribution (SED) fitting. The former make use of a large set of spectroscopic objects to calibrate some empirical relations between redshifts and photometry (i.e., photometric magnitudes and/or colors), e.g., Connolly et al.\ (1995) simply fit \zs~as a linear or quadratic function of magnitudes; Collister et al.\ (2004) developed their code ANNz based on an artificial neural network method. The empirical methods prove to be accurate; however they need a large number of training spectroscopic samples and potentially have large uncertainties for faint sources for which \zs~is sparsely sampled (Walcher et al.\ 2010). To evaluate the quality of \zp~derived by empirical methods more accurately and realistically, the authors often need to perform blind tests, i.e., randomly picking out a subsample of the \zs~sources for training and the rest for \zp~quality evaluation. The template SED fitting methods utilize library template sets and fit photometry at a series of redshift grid points to estimate \zp. Typically, no apparent training procedures are involved, the results are thus believed to be largely unbiased, but the choice of templates is essential in determining quality \zp\ estimates. 

There are additional factors such as resolution of photometric SED (depending on the number and bandwidths of filters) and wavelength coverage of filters that affect \zp~quality. A high-resolution photometric SED might capture some detailed spectroscopic features such as emission or absorption lines, thereby narrowing the possible redshift range. Wide wavelength coverage could eliminate the degeneracy of SEDs, thus reducing the probability of catastrophic failures. The largest uncertainty in \zp~estimation lies in photometric errors, especially for faint sources where the signal-to-noise ratio (S/N) is low (e.g., Dahlen et al.\ 2010, 2013). Furthermore, images of different bands are often obtained with different instruments and their point spread functions (PSFs) might differ significantly. Therefore the challenge is to measure the same fraction of light (i.e., accurate colors) for a source in images with different PSFs. The method of PSF matching has proven to be effective in obtaining uniform photometry across different instruments and filters, with large variations in PSFs and pixel scales taken into account (e.g., Cardamone et al. 2010; Dahlen et al. 2010). Another issue in photometry is blending. When two sources are close to each other (due to projection effects), photometry of one source might be contaminated by the light from the other source (e.g., Dahlen et al. 2010); on some occasions two such sources might even be detected as one single source if their angular separation is sufficiently small (i.e., comparable to the angular resolution of the observations).

\begin{figure*}
\includegraphics[width=\linewidth]{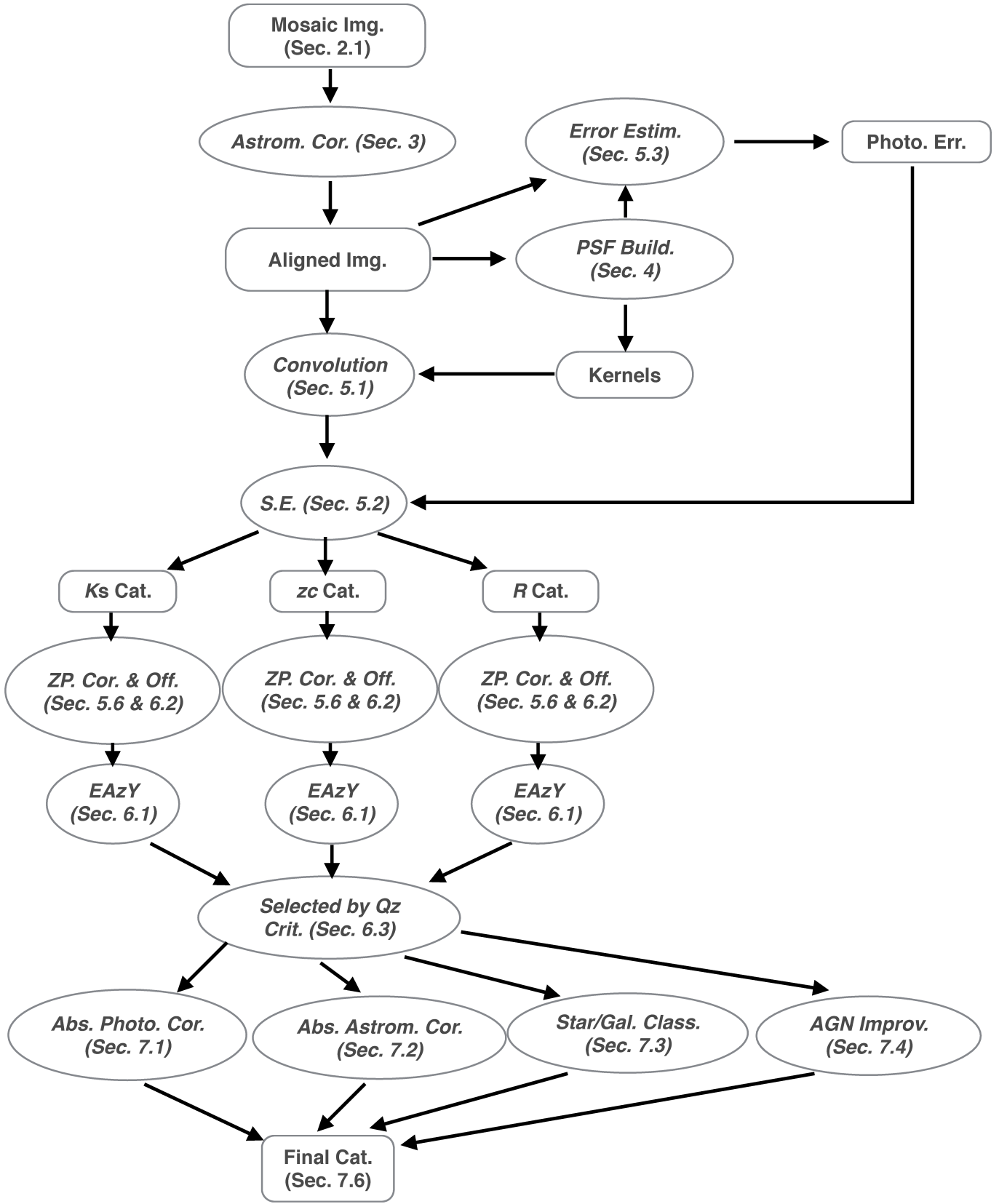}
\caption{The flow chart of our main procedures for \zp~estimation. For brevity we adopt the flowing abbreviations: Img.=Images, Astrom.=Astrometry, Off.=Offsets, Cor.=Corrections, Estim.=Estimation, Build.=Building, Photo.=Photometry, Err.=Error, S.E.=SExtractor, Cat.=Catalog, ZP.=Zero Point, Crit.=Criterion, Abs.=Absolute, Gal.=Galaxy, Class.=Classification, Improv.=Improvement, and Sec.=Section.}
\label{fig:flow}
\end{figure*}

In this paper, we perform PSF-matched photometry and determine photometric redshifts for over one hundred thousand objects in the Hawaii-Hubble Deep Field-North (\hdfn; Capak et al. 2004; C04 hereafter) that is an intensively-observed field. Centered at $\rm \alpha_{J2000.0}=12^h 37^m$ and $\rm \delta_{J2000.0}=+62\degree 10\arcmin$, the 0.4-deg$^2$ \hdfn\ contains the \goodsn\ (Giavalisco et al.\ 2004) and \cdfn\ (Brandt et al.\ 2001; Alexander et al. 2003, A03 hereafter) fields. Rafferty et al.\ (2011) calculated \zp~for the 48,858 sources in the C04 catalog, using photometry that is not PSF-matched and was collected from several origins. The recent \hbox{3D-{\it HST}} project (Skelton et al.\ 2014) derived \zp~based on deep \hst~data and some ground-based observations in the \goodsn\ field, which covers a much smaller area than the \hdfn. 
Given the fact that the \hdfn\ is a premium field with an enormous investment of multi-wavelength observations (in particular the recent addition of deep infrared data), it is imperative to produce a catalog that presents both PSF-matched photometry and photometric redshifts for the {\it entire} \hdfn~field. Therefore, we collect images in 17 broad bands from ultraviolet (\bandu~band, $\sim 0.3~\mu m$) to mid-infrared (IRAC $8.0\ \mu m$) and derive \zp~using the EAzY code (i.e., a SED fitting code developed by Brammer et al. 2008; B08 hereafter) based on 15 bands (excluding IRAC $5.8\ \mu$m and $8.0\ \mu$m; see Section~\ref{sec:zp}) in this field. Our main procedures are outlined in Figure~\ref{fig:flow}.

This paper is structured as follows. We describe the imaging and spectroscopic data in Section~\ref{sec:data}, astrometry correction in Section~\ref{sec:astrom}, PSF-matching procedures in Section~\ref{sec:psf}, and photometry extraction in Section~\ref{sec:photo}, respectively. In Section~\ref{sec:photoz}, we present the procedures used to derive \zp\ and evaluate our \zp~quality. We describe the corrections to obtain absolute photometry and astrometry, source classification, special treatment for \zp~of \xray~sources, advice on using our catalog, and catalog details in Section~\ref{sec:cat}. In Section~\ref{sec:sum}, we give a brief summary of this work. Throughout this paper, all apparent magnitudes are quoted in the AB system unless otherwise stated, where AB magnitude is defined as $\rm mag=23.9-2.5log(flux(\mu Jy))$. We assume a cosmology of $\Omega_M=0.3$, $\Omega_{\Lambda}=0.7$, and $H_0=70$~km~s$^{-1}$~Mpc$^{-1}$.

\section{data}\label{sec:data}

\subsection{Imaging Data}\label{sec:img}
\begin{figure}
\includegraphics[width=\linewidth]{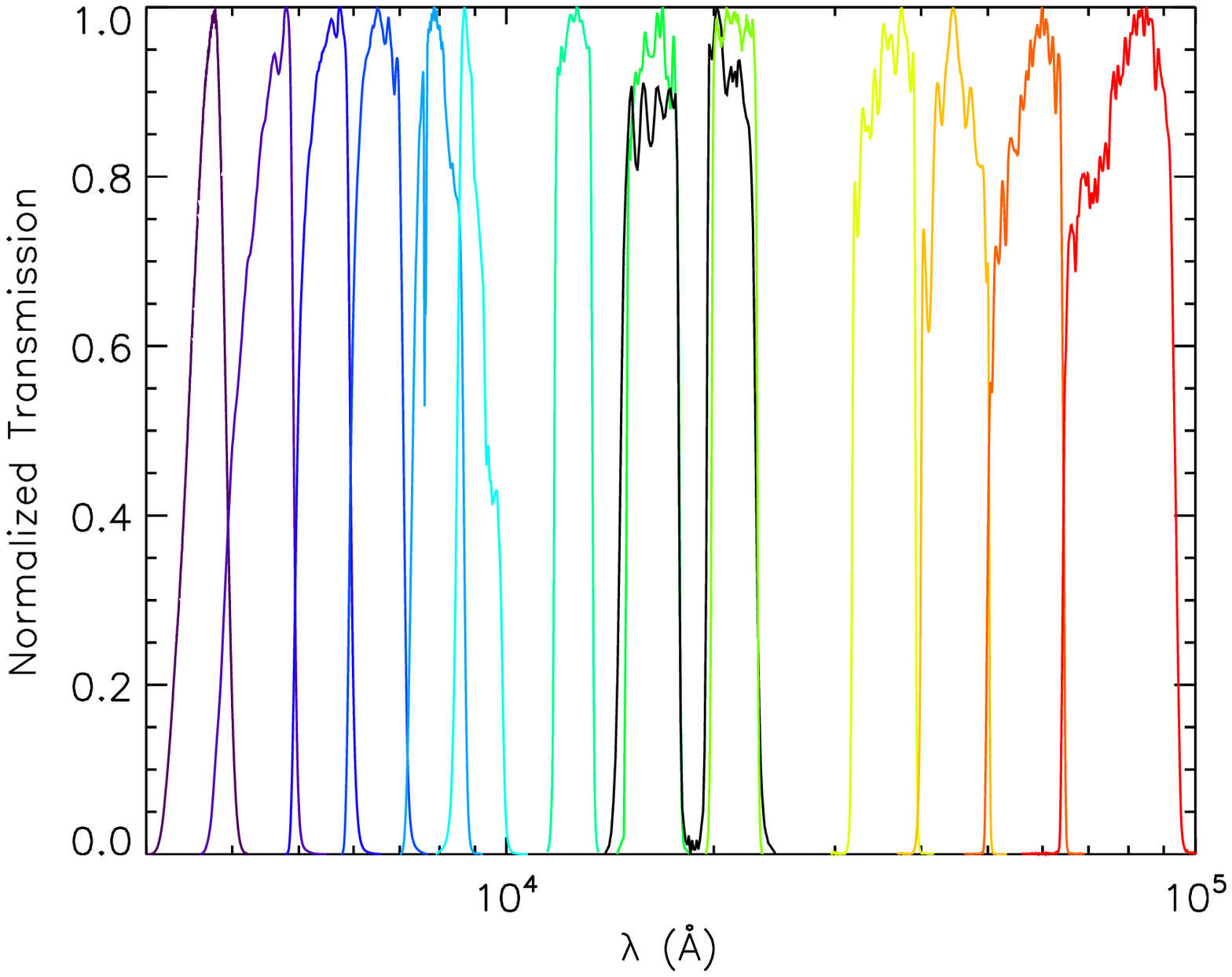}
\caption{Normalized filter transmission curves. From left to right (except \bandhk~that is indicated as a black curve), the curves are for the \bandu, \bandb, \bandv, \bandr, \bandi, \bandzc (\bandzo), \bandj, \bandh, \bandk, \sone (\aone), \stwo (\atwo), \sthr, and \sfor~bands, respectively.\\(A color version of this figure is available in the online journal.)}
\label{fig:filter}
\end{figure}

\begin{figure}
\includegraphics[width=\linewidth]{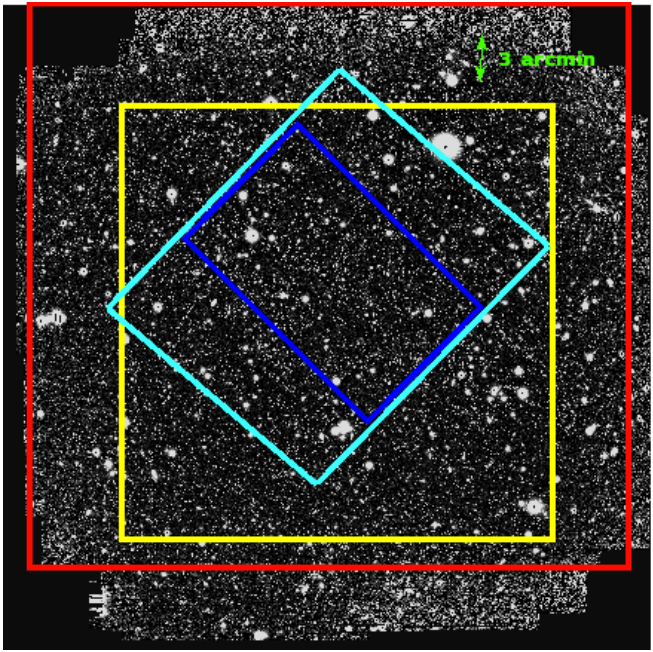}
\caption{\bandr-band image overlaid with rectangles indicating the \bandk~coverage (red), the field in C04 (i.e., the central \hdfn~field; yellow), the \cdfn~(cyan), and the \goodsn~region (blue), respectively. The \bandzc~band covers nearly the same region as the \bandr~band.\\(A color version of this figure is available in the online journal.)}
\label{fig:field}
\end{figure}

We collect the \bandu-, \bandb-, \bandv-, \bandr-, \bandi-, $z^\prime$-, and \bandhk-band~images from C04, the \bandj- and \bandh-band images~from Keenan et al.\ (2010), and the \bandk-band image~from Wang et al.\ (2010) (W10 hereafter), respectively. We also make use of an independently observed $z^\prime$-band image from Ouchi et al.\ (2009). The IRAC 3.6, 4.5, 5.8, and 8.0~$\mu$m images were obtained from the Spitzer Heritage Archive processed via Super-Mosaic pipeline version 2.0, calibration pipeline version S18.25.0, and MOPEX (for mosaic processing) version 18.5.4 (for 3.6, 4.5, and 5.8~$\mu$m) and 18.5.6a (for 8.0~$\mu$m), while another set of IRAC 3.6 and 4.5~$\mu$m images were taken from the Spitzer Extended Deep Survey (SEDS) presented in Ashby et al.\ (2013). The image information is listed in Table~\ref{tab:img} and the filter transmission curves are plotted in Figure~\ref{fig:filter}. We show the \bandr-band image overlaid with various coverages in the \hdfn~in Figure~\ref{fig:field}. The \goodsn, \cdfn, and \bandk-band coverages are encircled by the blue, cyan, and red rectangles, respectively. The yellow rectangle indicates the region (i.e., the central \hdfn~field) where C04 derived their catalogs. We find that, different from other images, all IRAC images have some residual background, which could bias the PSF
model building (see Section~\ref{sec:psf});
therefore, we run SExtractor to subtract the background before further analyses. 

\begin{table*}
\begin{center}
\caption{Imaging Data}\label{tab:img}\begin{tabular}{ccccccccr}\hline\hline
Band & Depth & PSF Size & Zero Point & Solid Angle & Group & Galactic Extinction & Epoch & Reference \\
(1) & (2) & (3) & (4) & (5) & (6) & (7) & (8) & (9)  \\ \hline
\bandu               & 26.3 & 1.63\arcsec & 31.369  & 0.42 & \two &      $-0.048$ &2002&      Capak et al.\ 2004\\
\bandb               & 26.3 & 1.13\arcsec & 31.136  & 0.31 & \one &      $-0.042$ &2001&      Capak et al.\ 2004\\
\bandv               & 25.8 & 1.56\arcsec & 34.707  & 0.39 & \two &      $-0.031$ &2001&      Capak et al.\ 2004 \\
\bandr               & 26.0 & 1.60\arcsec & 34.676  & 0.39 & \two &      $-0.025$ &2001&      Capak et al.\ 2004\\
\bandi               & 25.1 & 1.08\arcsec & 33.481  & 0.39 & \one & $-0.018$ &2001-2002& Capak et al.\ 2004\\
$z^\prime$ (\bandzc) & 24.9 & 1.08\arcsec & 33.946  & 0.39 & \one &      $-0.014$ &2000-2001& Capak et al.\ 2004 \\
$z^\prime$ (\bandzo) & 25.7 & 1.20\arcsec & 33.020  & 0.33 & \one &      $-0.014$ &2001-2007& Ouchi et al.\ 2009 \\
\bandj               & 24.5 & 1.11\arcsec & 23.900  & 0.22 & \one &      $-0.008$ &2006&      Keenan et al.\ 2010 \\
\bandh               & 22.9 & 1.32\arcsec & 23.900  & 0.36 & \one &      $-0.005$ &2008&      Keenan et al.\ 2010 \\
\bandk               & 23.7 & 1.08\arcsec & 23.900  & 0.36 & \one &      $-0.004$ &2006-2008& Wang et al.\ 2010 \\
\bandhk              & 22.3 & 1.20\arcsec & 30.132  & 0.11 & \one &      $-0.005$ &1999-2002& Capak et al.\ 2004 \\
$3.6\mu$m~(\aone)    & 25.1 & 2.53\arcsec & 21.581  & 0.33 & \thr &      $-0.002$ &2004-2011& Ashby et al.\ 2013 \\
$3.6\mu$m~(\sone)    & 24.5 & 2.40\arcsec & 21.581  & 0.33 & \thr &      $-0.002$ &2004-2006& Spitzer Archive\\
$4.5\mu$m~(\atwo)    & 24.6 & 2.53\arcsec & 21.581  & 0.33 & \thr &      $-0.002$ &2004-2011& Ashby et al.\ 2013 \\
$4.5\mu$m~(\stwo)    & 24.2 & 2.43\arcsec & 21.581  & 0.31 & \thr &      $-0.002$ &2004-2006& Spitzer Archive \\
$5.8\mu$m~(\sthr)    & 22.6 & 2.96\arcsec & 21.581  & 0.33 & \thr &      $-0.002$ &2004-2006& Spitzer Archive  \\
$8.0\mu$m~(\sfor)    & 22.7 & 3.24\arcsec & 21.581  & 0.28 & \thr &      $-0.002$ &2004-2006& Spitzer Archive  \\\hline
\end{tabular}
\end{center}
{\sc Note.} ---
Col.~(1): Band name.
Col.~(2): 5$\sigma$ limiting AB magnitude estimated with a 2.1\arcsec-diameter (7 pixels) aperture, based on background noise estimation detailed in Section~\ref{sec:phot err}.
Col.~(3): PSF size that is calculated based on the PSF models built in Section~\ref{sec:psf model}, which is defined as the aperture diameter that encircles half of the total flux.
Col.~(4): Zero point in units of AB magnitude.
Col.~(5): Solid-angle coverage in units of deg$^2$. We only consider the areas located in the \hdfn~field.
Col.~(6): Group name that is classified based on the PSF size of a band (see Section~\ref{sec:psf sol}).
Col.~(7): Galactic extinction in units of AB magnitude (see Section~\ref{sec:ext}).
Col.~(8): Observational epoch.
Col.~(9): Reference of imaging data. The Spitzer Archive data can be retrieved at \href{http://irsa.ipac.caltech.edu/applications/Spitzer/SHA/}{\textcolor{blue}{the Spitzer Heritage Archive}}.
\end{table*}

For convenience in describing the PSF-matching technique (see Section~\ref{sec:psf}), we denote PSF size as the aperture diameter that encircles half of the total flux. Hereafter, we refer to Ouchi's $z^\prime$ band as \bandzo~band, and Capak's $z^\prime$ band as \bandzc~band. The filters of \bandzo~and~\bandzc~are identical, but the images were taken during different observational epochs (C04; Ouchi et al.\ 2009). The 3.6~$\mu$m and 4.5~$\mu$m bands in Ashby et al.\ (2013) are referred to as \aone~and~\atwo~bands respectively, while the 3.6, 4.5, 5.8, and 8.0~$\mu$m bands from the Spitzer Heritage Archive are referred to as \sone, \stwo, \sthr, and \sfor~bands respectively. Filters of \aone~and~\sone, \atwo~and~\stwo~are identical, and the observational epochs of \sone~and~\stwo~were also used to derive the \aone~and~\atwo~images (Ashby et al.\ 2013; also see Table~\ref{tab:img}). Although the \sone~and \stwo~images are shallower, we still incorporate them in our \zp~estimation, because the deeper \aone~and~\atwo~images were obtained by stacking more observations and thus were potentially more blurred (see Table~\ref{tab:img} for a comparison between PSF sizes). Indeed, we find that the inclusion of the \sone~and~\stwo~bands improves slightly the \zp~quality (see Section~\ref{sec:zs cat} for the indicators of \zp\ quality). The application of the rest-frame template error function by EAzY gives much lower weights to mid-IR than optical bands, therefore the use of duplicate 3.6~$\mu$m and 4.5~$\mu$m bands should not affect the \zp~estimation significantly (see Section~\ref{sec:photoz}), as found above. We do not make use of the Hubble Space Telescope (\hst) data in our \zp~estimation for the following reasons. First, the \hst~images are restricted in the \goodsn~region, the area of which is only $\lsim$ 20\% of that of the \hdfn~(see Figure~\ref{fig:field}). Second, the \hst~images have much smaller PSF sizes than our images, but this great advantage would be compromised if we apply the same PSF-matched photometry-extraction procedures (see Section~\ref{sec:photo}) to them. Finally, the \hbox{3D-HST} team has derived photometry and photometric redshifts (Skelton et al.\ 2014) for the \goodsn~region to greater depths utilizing the \hst~data, and their \zp~catalog is available now and complements our catalog effectively. 

\subsection{\zs~Data}\label{sec:zs cat}
\begin{figure}
\includegraphics[width=\linewidth]{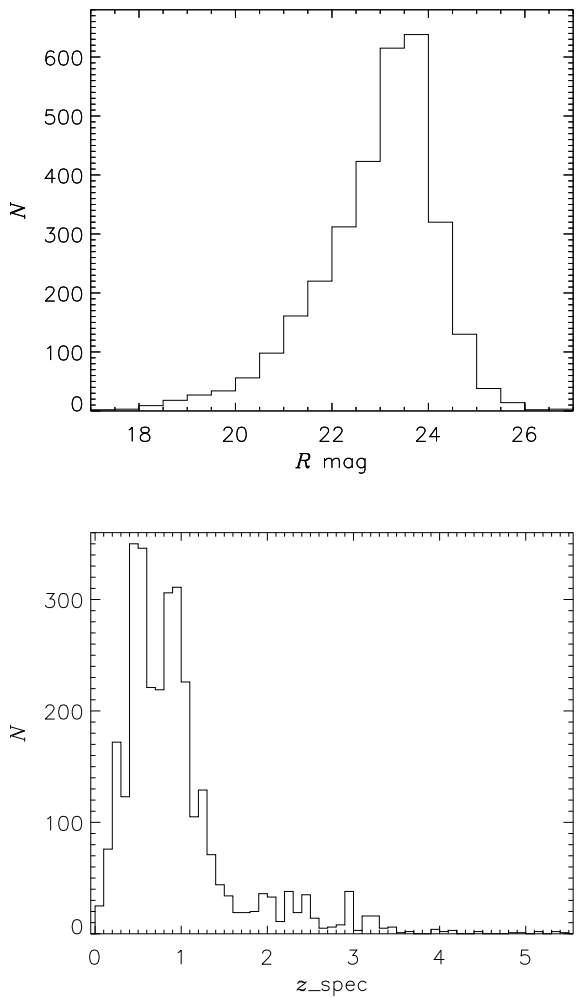}
\caption{Histograms of \bandr-band magnitude (Top) and \zs\ (Bottom) for all non-stellar \zs\ sources.}
\label{fig:spec_mag}
\end{figure}

The \zs\ data are collected from a number of references, with relevant information listed in Table~\ref{tab:zspec}. We only adopt secure \zs\ data that include at least two spectral features. Note that the \zs~data mainly come from Barger et al.\ (2008), which is due to the fact that a significant fraction of their data were compiled from previous works, e.g., Wirth et al.\ (2004). To remove duplicate \zs~entries, \zs~sources from different references are matched with each other using a 0.5\arcsec\ matching radius, except for those from Barger et al.\ (2003) that have already been matched by the authors to the 2~Ms {\it Chandra} Deep Field-North (CDF-N) main X-ray catalog (A03; see Section~\ref{sec:agn}). If one source has multiple \zs ~values and any two of those values are inconsistent (i.e.\ $|z_{ \rm{spec1} }-z_{\rm{spec2}}| / (1+z_{\rm{spec1}}) > 0.01$), we then discard all \zs~values of that source (less than 2\% of the \zs~values are discarded this way). If a source has only one \zs~value, we simply keep that \zs~for the source. Most of the \zs\ sources ($\sim 80\%$) are in the \goodsn\ region. In Figure~\ref{fig:spec_mag}, we plot in the top panel the \hbox{\bandr-band} magnitude (derived from this work; see Section~\ref{sec:photo}; upper limits not included) distribution of all non-stellar \zs~sources, which peaks around \bandr$=23.5$~mag and declines rapidly beyond that;
and we plot in the bottom panel the \zs\ distribution of all non-stellar \zs\ sources, the vast majority of which have \zs~$\lsim 1.6$.

We adopt the widely-used normalized median absolute deviation $\rm \sigma_{NMAD}$ (e.g., B08) to evaluate our \zp\ quality (see Section~\ref{sec:zp qua}), defined as
\begin{equation}\label{equ:sigma}
\rm \sigma_{NMAD}=1.48\times median(|\frac{\Delta \it \z - \rm median(\Delta \it \z)}{1+\it \z_{\rm spec}}|),
\end{equation}
where $\Delta z=\zp-\zs$.
Additionally,
we also examine outlier fractions of \zp\ results, with outliers being defined as sources having
$\rm |\Delta \z|/(1+\zs) >0.15$.

\begin{table*} 
\caption{\zs~references}\label{tab:zspec}
\begin{center}
\begin{tabular}{c r r r r} \hline \hline
Index & Reference & Non-X-ray & X-ray & Stars \\ 
(1) & (2) & (3) & (4) & (5) \\ \hline
1 & Barger, A. J. et al.\ (2008)& 2557 & 217 & 197 \\ 
2 & Cowie, L. L. et al.\ (2004)& 64 & 2 & 17 \\
3 & Reddy, N. A. et al.\ (2006)& 23 & 0 & 0 \\
4 & Wirth, G. D. et al.\ (2004)& 82 & 1 & 5 \\
5 & Cooper, M. C.  et al.\ (2011)& 111 & 3 & 2 \\
6 & Cohen, J. G. et al.\ (2000)&  5 & 0 & 1 \\
7 & Chapman, S. C. et al.\ (2005)& 3 & 1 & 0 \\ 
8 & Barger, A. J. et al.\ (2003)& 0 & 57 & 7 \\ \hline
 & Total & 2845 & 281 & 229 \\\hline
\end{tabular} \\
\end{center}
{\sc Note.} --- \
Col.~(1): Index of reference. If a source has consistent \zs~values in different references, then we attribute it to the reference with the lowest index.
Col.~(2): \zs~reference.
Cols.~(3--5): Numbers of additional unique \zs~sources that are non-\xray\ detected, \xray\ detected, and spectroscopic stars from each reference, respectively. The \xray\ source classification is based on matching with the A03 main \xray\ catalog (see Section~\ref{sec:agn}). 
If a source is classified as a star, then we do not classify it as an X-ray or non-X-ray source.
\end{table*}

\section{Astrometry Correction}\label{sec:astrom}
The above images are of various origins and thus have inconsistent astrometry, i.e., a source might have sightly different coordinates in different images. The systematic offsets among different images can be up to the order of $\approx0.3\arcsec$ in some areas. Given that Capak's 7 images (see Table~\ref{tab:img}) are well aligned with each other (C04), we adopt these images as standard, and align other images with them using {\sc geomap} and {\sc geotran} (IRAF\footnote{See \href{http://iraf.noao.edu/}{\textcolor{blue}{http://iraf.noao.edu/}.}} tasks). Specifically, we first run SExtractor (version 2.8.6; Bertin \& Arnouts~1996) on the \bandzc~image, whose wavelength is closer to the infrared bands than the other Capak bands, to locate standard objects, which are bright, not blended, unsaturated, and far from image borders. We then run SExtractor on the X (standing for \bandj, \bandh, \bandk, \bandzo, \aone, \sone, \atwo, \stwo, \sthr, and \sfor) band, and match the detected sources with those standard objects (typically there are $\sim$10,000 sources matched). {\sc geomap} uses these results to find an image manipulation solution (4th-order polynomial correction in our case, including linear manipulation such as rotation, shift, and rescaling as well as higher-order corrections), and {\sc geotran} executes the solution. We stress that we do not use stars exclusively as standard objects, because a fraction of registered stars are saturated in our deep images. Moreover, they are relatively sparse in our field, and thus the exclusive use of them might potentially compromise the astrometry in regions where no stars are present. 
Subsequently, we use hastrom.pro\footnote{See \href{http://idlastro.gsfc.nasa.gov/}{\textcolor{blue}{http://idlastro.gsfc.nasa.gov/}.}} (an IDL procedure) to transform all other images (i.e., \bandj, \bandh, \bandk, \bandzo, \aone, \sone, \atwo, \stwo, \sthr, and \sfor) to the same format as Capak's images, which are $8485\times8375$ arrays with a pixel size of 0.3\arcsec. This procedure enables performing photometry in the dual-image mode of SExtractor (see Section~\ref{sec:se}). 

\begin{figure}
\includegraphics[width=\linewidth]{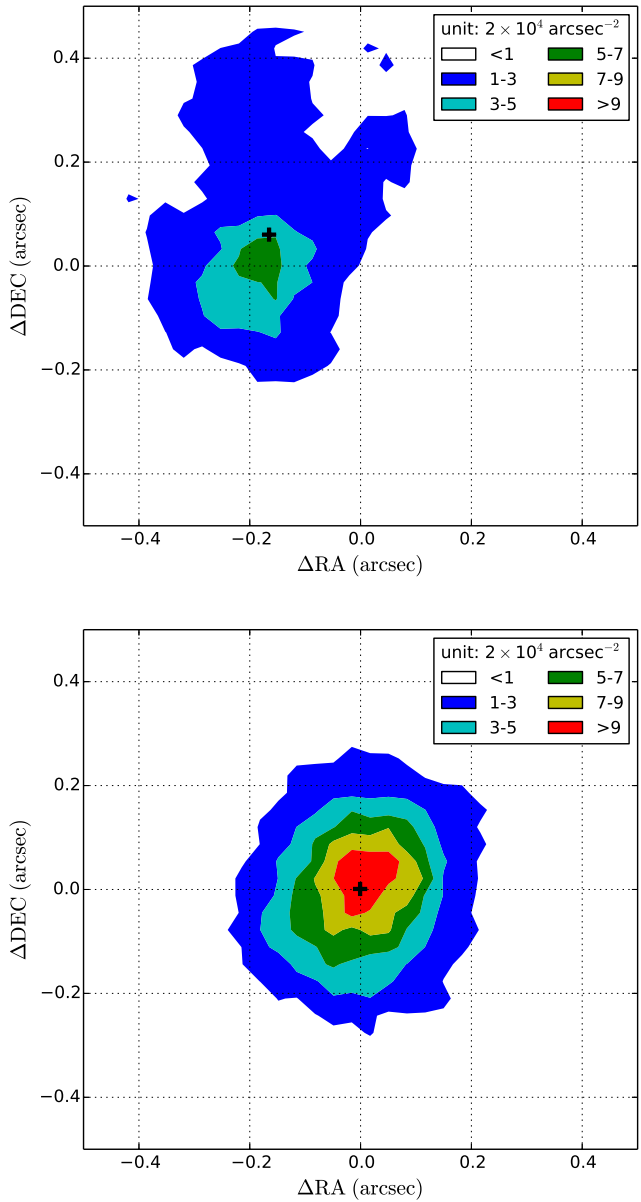}
\caption{Source density maps of coordinate offsets between the \bandk\ and \bandr\ bands. The top and bottom panels are based on data before and after astrometry correction. The black crosses indicate the median values of coordinate offsets. The contours represent different levels of source density, with color-coded scales shown in the insets. After the correction, the number of matched sources (with a matching radius of 0.5\arcsec) increases from 11,530 to 16,697 (i.e., a 44.8\% increase; note that the relatively small number of sources in this comparison is due to the application of much more stringent source-detection criteria than those presented in Section~\ref{sec:se}). \\(A color version of this figure is available in the online journal.)}
\label{fig:astrom}
\end{figure}

Figure~\ref{fig:astrom} plots the source density maps of coordinate offsets between the \bandk\ and \bandr\ bands before and after astrometry correction, which demonstrates the effectiveness of our procedures of astrometry correction.
In the construction of our final catalog (see Section~\ref{sec:abs astrom}),
we lock the absolute astrometry of our sources to the astrometry frame of the VLA data (Morrison et al. 2010), in order to facilitate cross matching between our catalog and other catalogs.
The reason why we adopt the astrometry frame of Capak's images, rather than that of the VLA data in the first place, is that we only care about relative astrometry when extracting photometry and want to make best use of the merit of the highly consistent astrometry among Capak's images. 

\section{Point Spread Function}\label{sec:psf}
As shown in Table~\ref{tab:img}, the image PSF sizes differ significantly, ranging from 1.08\arcsec\ (\bandi, \bandzc, and \bandk\ bands) to $\sim$ 3\arcsec~(IRAC bands). If a single-sized aperture were used to measure fluxes, it would encircle different fractions of light for images with different PSF sizes. Therefore it would be impossible to obtain accurate colors, which holds the key to obtaining accurate \zp\ measurements. One may propose to use different aperture sizes on different images, so that the same fraction of light could be captured. Indeed, this proposal may work for point-like sources whose profiles are simply PSFs of the images, given that we know the relation between the fraction of light encircled and the aperture size. However, for extended sources we do not know such relation due to the uncertainty in their shapes, thus this method is not practical. A common routine for consistent photometry is to smooth different images to the same PSF level before photometry extraction (e.g., Cardamone et al.\ 2010). In this way, two processed images could have identical PSFs theoretically, and apertures of same size will capture the same fraction of light. Below we describe our techniques to smooth one image to the PSF level of another.

 \subsection{PSF Models}\label{sec:psf model}
For each image, we match the detected sources with the stars in the GSC 2.3 catalog (Lasker et al.~2008), and use SExtractor's output flags to discard saturated, blended, and/or near-border ones. We then check source profiles, morphology, and contamination from nearby sources to further filter out surviving galaxies and saturated stars, and find that the profiles of the remaining stars resemble each other. About twenty stars in each image are selected this way to build the PSF models. We create a fixed-size thumbnail image centered on each star and normalize its flux to unity. We then construct the PSF model in the form of an image of the same size, whose pixel values are assigned as the median values of the corresponding pixels of those standard stars. Thus the PSF models represent the typical PSFs of corresponding images, with one single PSF model for each image.

 \subsection{PSF Smoothing}\label{sec:psf match}
\begin{figure}
\includegraphics[width=\linewidth]{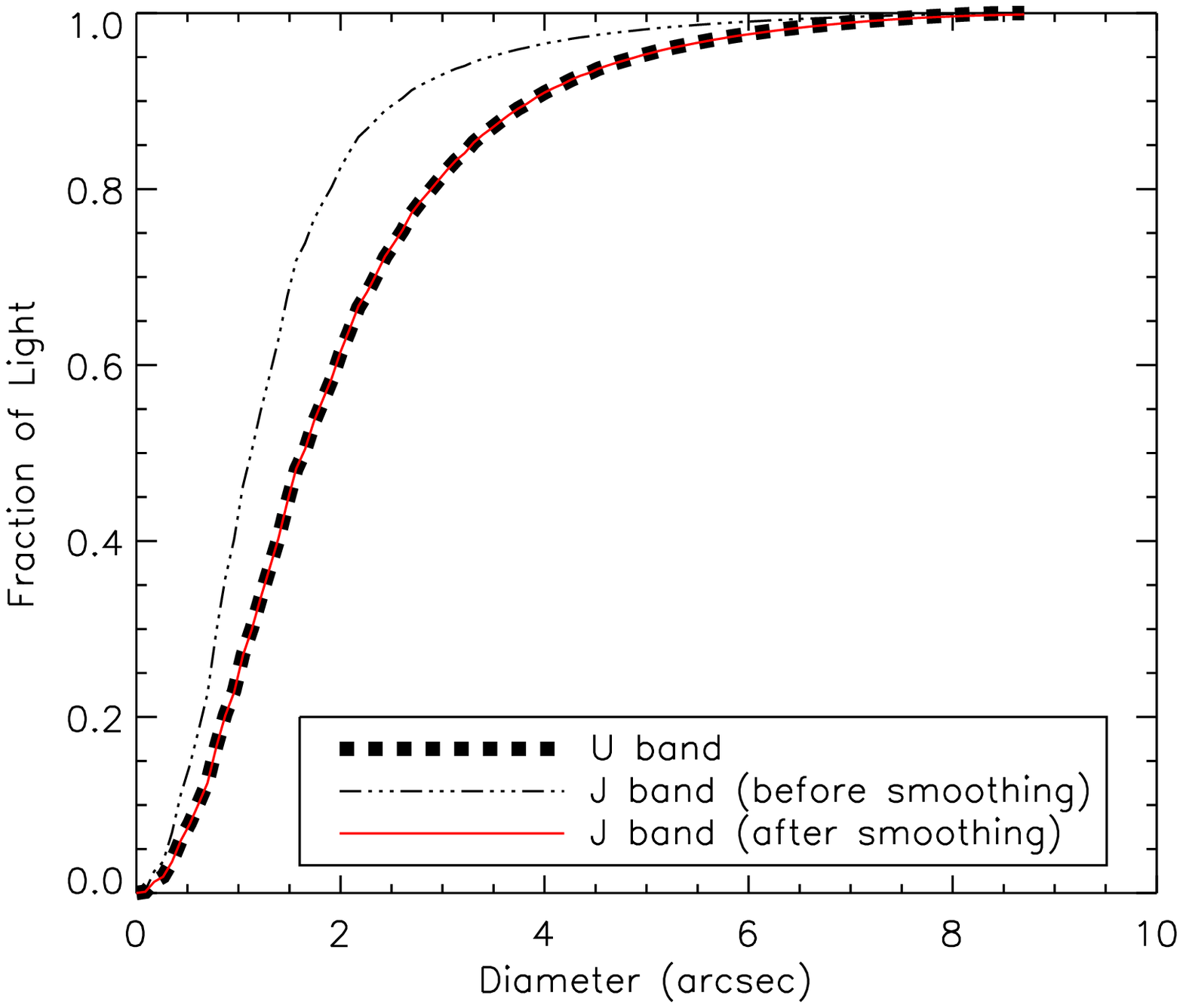}
\caption{Encircled fraction of light versus aperture diameter. Note that the curves of the \bandu~band and the smoothed \bandj~band are effectively identical.\\(A color version of this figure is available in the online journal.)}
\label{fig:psf}
\end{figure}

We make use of Lucy's procedure (IRAF task; Lucy 1974) to build the kernel to smooth one PSF to another PSF. Assuming there are two PSF thumbnail images, i.e., PSF A and PSF B as inputs (the corresponding images are denoted as image~A and image~B), Lucy's procedure will approximate the solution iteratively, which can then be used as the kernel to smooth image~A to the PSF level of image~B. The merit of Lucy's procedure is that it does not require that the PSFs are of specific modeled shapes such as Gaussian. It performs well when the target PSF size is much larger. In Figure~\ref{fig:psf}, we demonstrate the effectiveness of Lucy's procedure when smoothing the \bandj-band PSF (having the size of 1.11\arcsec) to the \bandu-band PSF (having a larger size of 1.63\arcsec). 

\section{Photometry Extraction}\label{sec:photo}
To conform with Capak's and Wang's catalogs (C04; W10), we construct three catalogs, i.e., \bandr-selected, \bandzc-selected, and \bandk-selected catalogs, respectively. The first two bands were used by C04 to detect sources, while the \bandk~band was used by W10. We perform aperture photometry via the dual-image mode in SExtractor (see Section~\ref{sec:se}). An appropriate choice of aperture size is critical in obtaining high-quality photometry: if the aperture is too big, then the S/N might be low leading to large contributions from noise; if it is too small, the uncertainty of astrometry might affect photometry significantly. We find that a choice of 1.5 times the PSF size as aperture diameter appears optimal for photometry extraction in terms of achieving good \zp~quality, which captures about 70\% of the light for a point-like source in each image. 

 \subsection{PSF Solutions}\label{sec:psf sol}
The PSF sizes can be divided into three groups (see Table~\ref{tab:img}):\\

\one. The \bandb-, \bandi-, \bandzc-, \bandzo-, \bandj-, \bandh-, \bandk-, and \bandhk-band images~have the smallest PSF sizes, i.e., $\lsim$ 1.3\arcsec; \\

\two. The \bandu-, \bandv-, and \bandr-band images~have moderate PSF sizes, i.e., $\approx$1.6\arcsec; \\

\thr. The \aone-, \atwo-, \sone-, \stwo-, \sthr-, and \hbox{\sfor-band} images~have the largest PSF sizes, i.e., $\gsim$ 2.4\arcsec.\\
 
If we smooth all images to the largest PSF, then the quality of the images with small PSF sizes, such as group \one~bands, will drop significantly, leading to much larger photometry errors. Therefore, we adopt the following approach to obtain accurate colors, and preserve the quality of the images maximally at the same time, which yields three catalogs (i.e., \bandr-, \bandzc-, and \bandk-selected) that are finally merged into one based on the $Q_z$ criterion (see Section~\ref{sec:qz}).

\subsubsection{\bandzc~and \bandk~Catalogs}\label{sec:ZcKs}
Here we describe how we perform PSF matching in order to obtain the \bandzc-~and \bandk-selected~catalogs.
For group \one~images, we smooth the \bandb-, \bandi-, \bandzc-, \bandzo-, \bandj-, \bandk-, and \bandhk-band images~to the PSF level of the \bandh-band image (the largest PSF size in group~\one); therefore the smoothed images are consistent with the \bandh-band image, and we extract the photometry of the detected sources. For the images in groups \two~and \thr, we adopt the strategy of aperture correction (e.g., Cardamone et al. 2010), which we describe below. First, we smooth image M (M stands for the detection band, i.e., \bandzc~or \bandk; see Section~\ref{sec:qz}) to image N (N stands for any band in groups \two~and \thr), and perform our photometry procedures on image~N and the smoothed image M. Then we define an aperture correction factor, cor$_{\textrm{aper}}$, for each detected source, as  
\begin{equation}\label{equ:cor_aper1}
\rm cor_{aper}=\frac{flux(M\ at\ PSF\ level\ of\ \bandh)} {flux(M\ at\ PSF\ level\ of\ N)},
\end{equation}
where flux is the total ADUs (Analog-to-Digital Units, i.e. the values of pixels in the image matrix) encircled by the aperture whose diameter equals 1.5 times the corresponding PSF sizes.
Finally we obtain the flux of N at the PSF level of the \bandh-band image for a source as 
\begin{equation}\label{equ:cor_aper2}
\rm flux(N)= flux(N\ not\ smoothed) \times cor_{aper}.
\end{equation}

\subsubsection{\bandr~Catalog}\label{sec:R}
The difference of the \bandr~band from the \bandzc~and \bandk~bands is that the \bandr~band belongs to group \two. Therefore, to obtain the $R$-band selected catalog, we cannot smooth the $R$-band image to the PSF levels of group \one~images, which have smaller PSF sizes. Thus we smooth all group~\one~and \two~images to the PSF level of $U$~band (the largest PSF size in group~\two), and apply aperture corrections to group \thr~images, following the procedure described in Section~\ref{sec:ZcKs}.

\subsection{SExtractor}\label{sec:se}
We use SExtractor to detect sources and extract photometry. At first, we detect many more sources in the \bandk-band image ($>$200,000 sources in total) than those in the \bandzc- and \bandr-band images (each having $<$100,000 sources) when using the same SExtractor parameters, but find that many of the \bandk-band sources are false detections through visual inspection. The reason is that, during the SExtractor runs, the \bandzc- and \bandr-band images are supplied with corresponding RMS maps, while the \bandk-band image with a weight map. Internally, SExtractor treats RMS maps as absolute noise levels, while weight maps are treated as relative noise levels; it then scales weight maps to absolute RMS maps via an internal algorithm. However, this algorithm tends to underestimate the noise level. To avoid the vast majority of false detections, W10 adopted the most stringent cleaning procedure, i.e., setting CLEAN\_PARAM = 0.1.\footnote{To avoid false detections due to bright objects, SExtractor assumes Moffat profiles for bright sources, and subtracts these profiles in the image to see if their faint neighbors could have been detected. This is the so-called `clean' procedure, and CLEAN\_PARAM controls the shape of the Moffat profile.} This configuration reduces effectively the number of \bandk-band detected sources to $\approx$90,000. Visual inspection shows that false detections are rare in this case. The other parameters in W10 were also chosen carefully and are therefore reliable, and, consequently, we adopt most of their parameters and run SExtractor on the dual-image mode. Table~\ref{tab:se} lists the main SExtractor parameters adopted for the \bandk-band image. There are only minor changes in SExtractor parameters for other images. Specifically, we lower the detection thresholds by setting DETECT\_MINAREA=2 and CLEAN\_PARAM=1 for the \bandzc\ and \bandr\ bands due to the difference between using their RMS maps and using the \bandk-band weight map as stated above. Generally these detection thresholds are sufficiently low to detect very faint sources. However, this might potentially lead to some false detections. Users of our catalog should be aware of this issue (see Section~\ref{sec:guide}).

\begin{table}
\caption{Main Parameters of SExtractor}\label{tab:se}
\begin{center}
\begin{tabular}{ l p{3cm} } \hline \hline
DETECT\_MINAREA  & 4 \\  
THRESH\_TYPE & relative \\
DETECT\_THRESH & 1.25 \\
FILTER & Y \\
FILTER\_NAME & gauss\_1.5\_3x3.conv \\
DEBLEND\_NTHRESH & 64 \\
DEBLEND\_MINCONT & 0.00001 \\
CLEAN & Y \\
CLEAN\_PARAM & 0.1 \\
MASK\_TYPE & correct \\
PHOTO\_APERTUES & 6.62, 8.16, 7.79, 8.02, 12.66, 12.01, 12.15, 14.79, 16.18 \\
PHOTO\_AUTOPARAMS & 2.5, 3.5 \\
GAIN & 0$^a$ \\
PIXEL\_SCALE & 0.3 \\
BACK\_TYPE & auto \\
BACK\_SIZE & 32 \\
BACK\_FILTERSIZE & 6 \\
BACKPHOTO\_TYPE & local \\
BACKPHOTO\_THICK & 24 \\
BACK\_FILTTHRESH & 0.0 \\ \hline
\end{tabular}
\end{center}
{\sc Note.} --- 
$^a$ In SExtractor, gain is only used in calculating flux uncertainty. Setting gain=0 means gain=Infinity, i.e., not including Poisson errors. There are two reasons for this: 1.\ Many images do not contain the gain values in their headers, and we do not find the information in related papers either. 2. Poisson errors are often small compared to background noise and can be neglected. W10 also set the gain to 0.
\end{table}

\subsection{{Photometric Errors}}\label{sec:phot err}
SExtractor assumes pixel-uncorrelated errors, and neglects background noise contributed by faint sources below the detection threshold. Therefore its derived photometric errors are underestimated. As pointed out by, e.g., Dahlen et al.\ (2013), accurate photometric errors are crucial in deriving accurate \zp. To obtain a more accurate background noise estimate for each source, we place around the source 100 apertures of the same size used for photometry extraction. These apertures are placed avoiding sources shown in the SExtractor-provided segmentation checkimage, and do not overlap with each other. Then we calculate the standard deviation of the flux in those apertures, and use it to replace the error given by SExtractor. The new errors, typically being several times larger than those provided by SExtractor, result in much better \zp~quality. Furthermore we also consider additional errors introduced by our PSF-matching procedure. We thus introduce $\rm err_{psf}$: if we smooth band P to the PSF level of band Q, then $\rm err_{psf}$ is defined as 
\begin{equation}\label{equ:phot err}
\rm err_{psf}=\frac{ |FL(smoothed\ P) - FL(Q)| }{FL(Q)},
\end{equation}
where FL means the fraction of light encircled by the photometry aperture in the PSF image. 
$\rm err_{psf}$ is a relative quantity, and is multiplied by the flux before being added quadratically to the aforementioned new error. For most bands, $\rm err_{psf}$ is $\lsim 4\%$. Typically, the background error dominates the $\rm err_{psf}$. For those objects whose flux in one band is less than the corresponding error, an upper limit is assigned by setting the flux to the value of the error and is included in the \zp~derivation (see Section~\ref{sec:photoz}). 
 
 \subsection{Photometric Consistency}\label{sec:photo con}
We note that one photometric band might have two images (e.g., \bandzc~and \bandzo~bands). If large photometric differences exist between the two images, it will be impossible to fit both sets of photometry well with templates in the \zp~estimation. We find that even a single inconsistent band could often ruin \zp~quality, regardless of how perfect the other bands are. To reduce inconsistency, we first eliminate the systematic offset ($\lsim 0.03$~mag) between the two fluxes by adjusting the zero point of either band in order to meet the condition of median(mag1$-$mag2)=0. 
This adjustment is adopted only to facilitate discarding inconsistent photometry and is not applied to the photometry for \zp\ estimation.
Then, for each source, if
\begin{equation}\label{equ:photo con}
\rm |mag1-mag2| > 3\times max(err\_mag1, err\_mag2),
\end{equation}
both sets of photometry will be discarded ($\lsim 5\%$ of sources have inconsistent fluxes in at least one band). The main reasons for inconsistent photometry are blending effects in crowded fields and contamination from nearby bright sources.

 \subsection{Galactic Extinction Correction}\label{sec:ext}
The \hdfn~field, located at high Galactic latitude, was initially chosen to be subject to minimal Galactic extinction. We apply Galactic extinction corrections obtained from an online utility provided through the NASA/IPAC Infrared Science Archive.\footnote{See \href{http://irsa.ipac.caltech.edu/applications/DUST/}{\textcolor{blue}{http://irsa.ipac.caltech.edu/applications/DUST/}} for details.} As shown in Table~\ref{tab:img}, the corrections, as expected, are small, ranging from 0.002 mag for the \aone, \atwo, \sone, \stwo, \sthr, and \sfor~bands to 0.048 mag for the \bandu~band.

 \subsection{Zero-Point Corrections}\label{sec:zp cor}
The above procedures such as PSF matching and aperture correction are all likely to introduce systematic errors into the photometry. 
We derive zero-point corrections to account for this factor by fitting the photometry for spectroscopic stars (see Section~\ref{sec:zs cat}) with a set of 235 stellar templates at $z=0$.
The stellar templates are taken from the $Le~Phare$ photometric redshift and simulation package (Arnouts et al.\ 1999; Ilbert et al.\ 2006), which consist of 231 stellar spectra that include all normal spectral types and luminosity classes (Pickles 1998; Chabrier et al.\ 2000) and 4 white dwarf spectra (Bohlin, Colina \& Finley 1995).
Here we only consider the bands bluer than the IRAC ones because the stellar templates of Pickles (1998) do not cover wavelengths beyond 2.5~$\mu$m. 

We first tentatively match our sources with the \zs\ catalogs (see Section~\ref{sec:zs cat} and Figure~\ref{fig:spec_mag}) using a 0.5\arcsec\ matching radius and then remove the systematic astrometry difference ($\approx 0.08\arcsec$) before matching them again. 
The false-matching rate is estimated to be $\lsim 3\%$ by systematically shifting source coordinates and then recorrelating them.
We then obtain the new zero points as follows, using a total of $\approx 130$ spectroscopic stars that are not saturated or blended based on visual inspection,\footnote{We do not calculate the zero points iteratively because of the limited number of qualified spectroscopic stars. In fact, the results do not converge even after more than 10 times of iterations.}
\begin{equation}\label{equ:zp}
\rm new\_zp=old\_zp - 2.5\times log (median(\frac{fitting\_flux}{observed\_flux})).
\end{equation}
The results are listed in Table~\ref{tab:zp cor}. Note that for the three detection bands (i.e., \bandk, \bandzc\ and \bandr), the zero-point corrections are not exactly the same. This should be due to the fact that the photometry from different catalogs is derived from different sets of images (see Section~\ref{sec:psf sol}) and there are some uncertainties associated with the kernels used in smoothing (see Section~\ref{sec:phot err}) when obtaining those images. 

\begin{table} 
\begin{center}
\caption{Zero-Point Corrections (unit: mag)}\label{tab:zp cor}
\begin{tabular}{ c r r r} \hline \hline
          &\bandk      &\bandzc     &\bandr  \\ \hline
\bandu    &    0.227   &    0.190   &    0.193  \\
\bandb    & $-$0.024   & $-$0.011   & $-$0.015  \\
\bandv    & $-$0.102   & $-$0.125   & $-$0.119 \\
\bandr    &    0.026   &    0.014   &    0.009 \\
\bandi    &    0.023   &    0.035   &    0.016 \\
\bandzc   & $-$0.043   & $-$0.029   & $-$0.022 \\
\bandzo   &    0.114   &    0.110   &    0.113  \\
\bandj    & $-$0.025   & $-$0.032   & $-$0.036  \\
\bandh    &    0.039   &    0.045   &    0.044  \\
\bandk    &    0.025   &    0.018   &    0.032 \\
\bandhk   & $-$0.181   & $-$0.154   & $-$0.056  \\ \hline
\end{tabular} \\
\end{center}
{\sc Note.} --- 
Row names indicate photometry bands, and column names indicate source-detection bands. The zero points of IRAC bands are not corrected because of limited wavelength coverage of our stellar templates.\\
\end{table}

\begin{figure*}
\includegraphics[width=\linewidth]{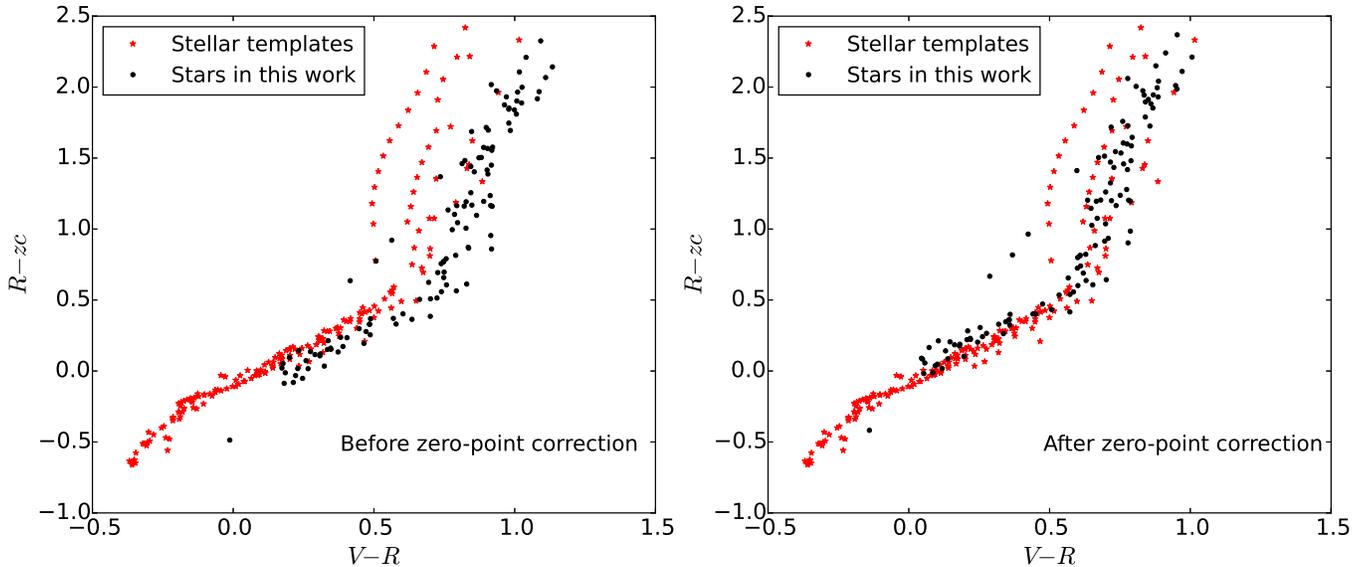}
\caption{\bandv\bandr\bandzc\ color-color plots for the spectroscopic stars in our catalog (black bullets; only those with photometry errors less than 0.1~mag in all the three bands are plotted) and the model stars derived with stellar templates (red stars).
(Left) The case without correcting for zero points for our spectroscopic stars.
(Right) The case with zero-point corrections applied to our spectroscopic stars.\\(A color version of this figure is available in the online journal.)}
\label{fig:star_phot}
\end{figure*}

We plot \bandv\bandr\bandzc\ color-color plots in Figure~\ref{fig:star_phot} to show the effect of zero-point corrections. For the case without correcting for zero points (i.e., the left panel of Figure~\ref{fig:star_phot}), overall, our star colors do not agree well with template star colors. However, for the case with zero-point corrections applied (i.e., the right panel of Figure~\ref{fig:star_phot}), our star colors are in good agreement with template star colors.
Therefore, we conclude that zero-point correction is effective in eliminating systematic errors and adopt the new zero points (i.e., the corrected photometry) for subsequent analyses.

\subsection{Completeness}\label{sec:complete}

Our source-detection approach recovers 46,914 sources out of a total of 48,858 sources with $\ge 5\sigma$ significance in the C04 catalog using a 0.5\arcsec~matching radius after removing any systematic astrometry offsets (46914/48858=96\%). For the \bandk-band selected catalog of W10, the recovered number of sources is 53,544 out of a total of 56,967 sources with $\ge 5\sigma$ significance (53544/56967=94\%). Through visual inspection, we find that nearly all the unmatched sources are very faint or blended with other sources, and thus source detection and position determination are more uncertain.

\begin{figure}
\includegraphics[width=\linewidth]{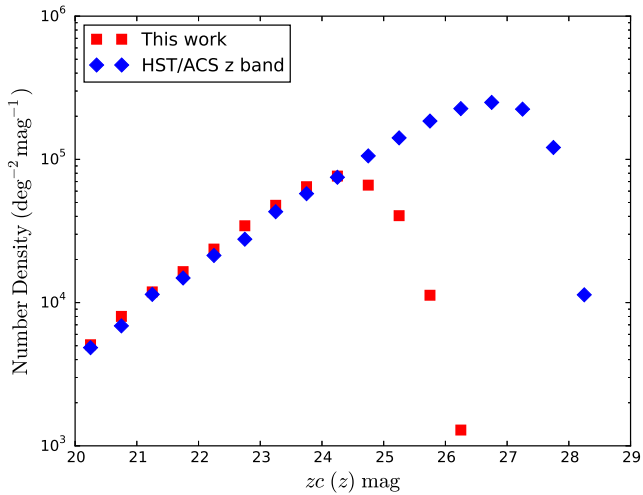}
\caption{Source density (number counts) plots for our \bandzc\ band (red squares) and the \hst\ \textit{z}(F850lp) band of the \goodsn\ catalog (blue diamonds).
Only sources with $\ge 3\sigma$ significance are counted. For direct comparison, we only consider our \bandzc-band sources that are located within the \goodsn\ region. The \goodsn\ \textit{z}(F850lp)-band magnitudes are the MAG\_AUTO magnitudes in SExtractor that are corrected by the same correction factor as mentioned in Section~\ref{sec:abs phot}.\\(A color version of this figure is available in the online journal.)}\label{fig:completeness}
\end{figure}

To estimate the completeness level of our catalog, we compare our magnitude-dependent source density with that of the \goodsn\ catalog (Giavalisco et al.\ 2004), and plot the source density histograms for our \bandzc\ band and the \hst\ \textit{z}(F850lp) band of the \goodsn\ catalog in Figure~\ref{fig:completeness}. For both catalogs, we only count sources with $\ge 3\sigma$ significance. Assuming that the \goodsn\ \textit{z}(F850lp) catalog is complete at least down to \textit{z}$\approx 25.5$~mag, our catalog is then $\approx 100\%$ and $\approx 80\%$ complete down to \textit{z}$\approx 24.5$~mag and 25.5~mag, respectively. Note that in some magnitude ($\lsim 24$~mag) bins, our number counts appears slightly higher than that of \hst\ \textit{z}(F850lp) band. This might be due to differences in filters and instruments.

 \subsection{Blending}\label{sec:blend}
When detecting sources, SExtractor might fail to separate two very close sources. 
To estimate the importance of this effect in our catalog, we make use of the \goodsn\ catalog (Giavalisco et al.\ 2004) that is based on \hst\ observations with superb angular resolutions of $\approx 0.05\arcsec$. 
We only estimate the effect for our \bandr~band, the one with the largest PSF size among the three detection bands, to assess an approximate upper limit of this effect. 
We then only consider sources with \bandr~$\lsim$26~mag (i.e., $5\sigma$ detection limit of our \bandr\ band; see Table~\ref{tab:img}).
Through visual inspection of both our \bandr-band and the \hst\ \textit{F606W}-band images (the \hst\ \textit{F606W} band is similar to our \bandr~band and both bands are centered at $\approx 6000\AA$), we find that the detection generally fails if the angular separation of two sources is less than 
1.2\arcsec. 
We also find that $\approx 10.8\%$ of the \hst\ \textit{F606W}$\lsim$26~mag sources in the \goodsn\ catalog  have neighboring sources within a radius of 1.2\arcsec.  
We therefore conclude that about $10.8\%/2=5.4\%$ of our \bandr-band sources might be in fact two sources that can only be separated distinctly in images with higher angular resolution. 

Another issue caused by blending is photometry contamination by nearby sources. 
We assume that photometry of one source is subject to contamination if another source is present within a radius of 1.5$\times$PSF size. 
For non-IRAC bands, we estimate the effect for our \bandu~band, the one with the largest PSF size of 1.63\arcsec, to assess an approximate upper limit of this effect. 
We find that $\approx$28\% of the \bandu-band sources might suffer from such photometry contamination. 
However, the actual contamination effect should be much less severe and could be totally negligible in many cases.
For example, bright sources are minimally affected (if at all) by their faint neighbors.
Furthermore, SExtractor can reduce this contamination effect when computing photometry (see Page~40 of the SExtractor manual for version 2.13). 
For IRAC bands, the photometry contamination is expected to be worse considering their large PSF sizes. 
However, this situation is largely alleviated given that IRAC data have much lower weights than optical and near-infrared data in \zp\ estimation (see Section~\ref{sec:img}).

Sources that suffer from the above blending issues would generally have inaccurate photometry and thus \zp\ of poor quality.
Such sources typically have large $Q_z$ (the redshift quality parameter defined in Section~\ref{sec:qz}) values, e.g.,
for sources with $\ge 5\sigma$ significance, $\approx 56\%$ of the (likely) blended sources have $Q_z>1$, in contrast to $\approx 23\%$ of the non-blended sources having $Q_z>1$.
In such cases, their \zp\ are generally not recommended for use (see Sections~\ref{sec:qz} and \ref{sec:guide}).

 \section{Photometric Redshifts}\label{sec:photoz}

\subsection{EAzY}\label{sec:eazy}
We use the EAzY code (B08) to estimate photometric redshifts. EAzY (version 1.00) can fit with linear combinations of template SEDs. There are two novel features of EAzY. First, the default template set (see Figure~\ref{fig:tem}) is based on semi-analytical models rather than spectroscopic samples (usually these are highly biased). B08 used the ``nonnegative matrix factorization'' algorithm to extract a template set from the library of $\sim$3000 P\'EGASE models (Fioc \& Rocca-Volmerange 1997) with ages ranging from 1~Myr to 20~Gyr and having a variety of star formation histories. Second, B08 designed a rest-frame template error function to estimate template uncertainty. This function assigns different wavelength regimes different weights (being largest, moderate, smallest for the rest-frame optical, ultraviolet, and near-infrared bands, respectively), and ensures that the formal redshift uncertainties are realistic (B08). 

\subsubsection{Templates}\label{sec:tem}
\begin{figure}
\includegraphics[width=\linewidth]{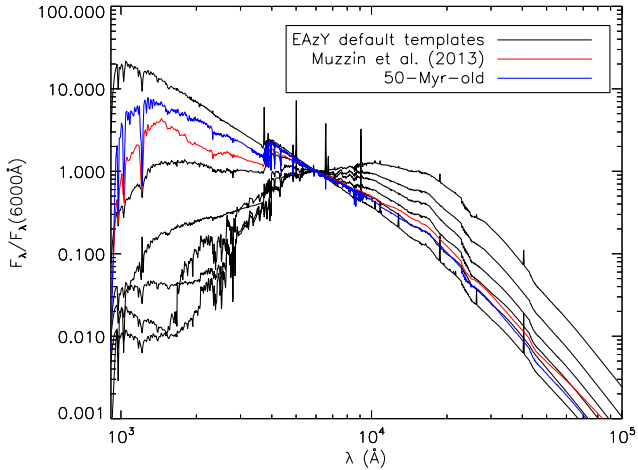}
\caption{Galaxy templates adopted in this work, including the six EAzY v1.00 default templates (black curves), the one taken from Muzzin et al.~(2013) (red curve), and a 50-Myr-old galaxy template generated by GALAXEV (blue curve).\\(A color version of this figure is available in the online journal.)}
\label{fig:tem}
\end{figure}

We find that the default template set of EAzY does not represent young galaxies well,
leading to large \zp\ uncertainties for such galaxies. 
To obtain more accurate \zp~for this population, we introduce two additional representative young galaxy templates (see Figure~\ref{fig:tem} for a total of eight galaxy templates adopted). The first one is a lightly dust-reddened young galaxy taken from Muzzin et al.\ (2013), and is designed to describe the most massive subset of the Lyman break galaxy population. This template improves the overall quality of \zp. The other template is a 50-Myr-old single-burst model with metallicity $\rm Z = Z_\sun$ generated by GALAXEV (Bruzual and Charlot 2003). This model improves the \zp~quality of objects at $z\gsim2$. 

To examine the effect of degeneracy introduced by adding these two young galaxy templates, we compare \zp\ results obtained for the sources detected in the \bandr~band with $>5\sigma$ significance, using the default template set of EAzY and using all eight galaxy templates shown in Figure~\ref{fig:tem}, respectively. We find that the best-fit templates and thus \zp~values do not change at all for the majority of the sources when these two new templates are introduced, thereby resulting in a nominal \sigm~$=0.000$; and we obtain a nominal outlier fraction of 2.6\%. Therefore, the effect of this degeneracy is negligible.
  
\subsubsection{Other Parameters}\label{sec:oth}
We adopt the default mode of linear combination of templates and the featured template error function when fitting the photometry. The estimation of intergalactic medium (IGM) absorption is based on the default IGM absorption law (Madau 1995) in EAzY. We do not apply apparent magnitude priors when using EAzY, because we find this configuration tends to underestimate redshifts especially for high-redshift galaxies. We set the redshift grid from 0.01 to 8.0 with a logarithmic step size of $\Delta {\rm ln}(1+z)=0.01$. For each source, we adopt the \z~grid value that minimizes the fitting $\chi^2$ as our \zp~value.  

\subsection{Zero-Point Offsets}\label{sec:zp}
In Section~\ref{sec:zp cor}, we correct our photometry based on SED fitting of spectroscopic stars. However, the corrected photometry might still have systematic differences compared to the expected photometry based on galaxy templates that are used in \zp\ calculation. In order to achieve better \zp\ quality, we derive zero-point offsets to eliminate such photometry differences by fitting the photometry at \zs\ (see Section~\ref{sec:zp cor} for details of matching our sources with the \zs\ catalogs) with the eight adopted templates (see Figure~\ref{fig:tem}). 
We compute the zero-point offsets by
\begin{equation}\label{equ:zp off}
\rm zp\_offset = - 2.5\times log (median(\frac{fitting\_flux}{observed\_flux})).
\end{equation} 
After three iterations, the zero-point offsets converge (vary by $\lsim$0.01~mag). The zero-point offsets are listed in Table~\ref{tab:zp}.

\begin{table} 
\begin{center}
\caption{Zero-Point Offsets (unit: mag)}\label{tab:zp}
\begin{tabular}{ c r r r} \hline \hline
          &\bandk      &\bandzc     &\bandr  \\ \hline
\bandu    & $-$0.147   & $-$0.135   & $-$0.130 \\
\bandb    & $-$0.007   & $-$0.001   &    0.009 \\
\bandv    &    0.029   &    0.022   &    0.028 \\
\bandr    &    0.048   &    0.034   &    0.035 \\
\bandi    & $-$0.034   & $-$0.027   & $-$0.021 \\
\bandzc   & $-$0.022   & $-$0.017   & $-$0.009\\
\bandzo   & $-$0.005   &    0.002   & $-$0.001 \\
\bandj    &    0.020   &    0.031   &    0.016 \\
\bandh    &    0.009   &    0.017   & $-$0.012 \\
\bandk    &    0.104   &    0.123   &    0.101\\
\bandhk   &    0.074   &    0.082   &    0.043  \\
\aone     & $-$0.033   & $-$0.055   & $-$0.042  \\
\sone     & $-$0.045   & $-$0.064   & $-$0.048  \\
\atwo     & $-$0.035   & $-$0.056   & $-$0.037  \\
\stwo     & $-$0.028   & $-$0.042   & $-$0.025 \\
\sthr     &    0.096   &    0.075   &    0.101 \\
\sfor     &    0.367   &    0.351   &    0.378 \\ \hline
\end{tabular} \\
\end{center}
{\sc Note.} --- 
Row names indicate photometry bands, and column names indicate source-detection bands. \\
\end{table}

The zero-point offsets for the \sthr~and \sfor~bands indicate that our photometry has fluxes higher than template fluxes at $>5\mu$m~wavelengths. However, the \hbox{$\sthr - \sfor$} color~of stars without zero-point offsets applied appears more consistent with that of Stern et al.\ (2005). Furthermore, the inclusion of these two offset bands in our \zp~calculation yields a worse \zp~quality (see Section~\ref{sec:zp qua}). Therefore we conclude that the zero-point offsets for the \sthr~and \sfor~bands are unreliable, likely due to the absence of PAH emission features in our templates. We thus discard zero-point offsets for these two bands and do not use their photometry for \zp~estimation; however, for completeness, we still provide their photometry in the final catalog (see Section~\ref{sec:cat}). 

We find that our star colors generally become inconsistent with template star colors after applying the above zero-point offsets that are derived with galaxy templates. Such zero-point offsets are apparently template-dependent. Therefore, we adopt these zero-point offsets only in \zp\ estimation, but do not apply them to the photometry presented in the final catalog (see Section~\ref{sec:catdet}).

\subsection{\bandk, \bandzc, or \bandr~Catalog?}\label{sec:qz}
As described above, we have three catalogs derived from detections in the \bandk-, \bandzc-, and \bandr-band images, respectively. 
When merging the \bandk-, \bandzc-, and \hbox{\bandr-band} catalogs,
we use a matching radius of 1.0\arcsec\ because
the positions of a same source in different detection images could sometimes have
shifts exceeding 0.5\arcsec\ even after astrometry correction, due to
the fact that such sources (a total of $\approx 5000$, i.e., $\lsim 4\%$ of the final catalog; see the latter part of this section) are typically very faint and/or blended to some degree in certain bands and thus have inaccurate positions therein. 
About 60\% of the sources are detected in more than one band; their \zp\ values derived from different catalogs are generally very similar even though we use different approaches to produce the three catalogs. 
However, in some cases these sources do have apparently different \zp\ values from different catalogs because the \zp~qualities for given same source present in all three catalogs might be different: (1) if a source appears faint in one detection image, then its position and aperture correction factor (see Section~\ref{sec:psf sol}) are likely to be determined relatively poorly in that image, which might further lead to large uncertainties in photometry; and (2) different catalogs are in fact obtained from differently processed image sets (see Section~\ref{sec:psf sol}), and technically speaking, noise levels and blending conditions differ among those image sets. Therefore, proper choices among the three catalogs should enhance the overall \zp~quality. 

\begin{figure*}
\center
\includegraphics[width=\linewidth]{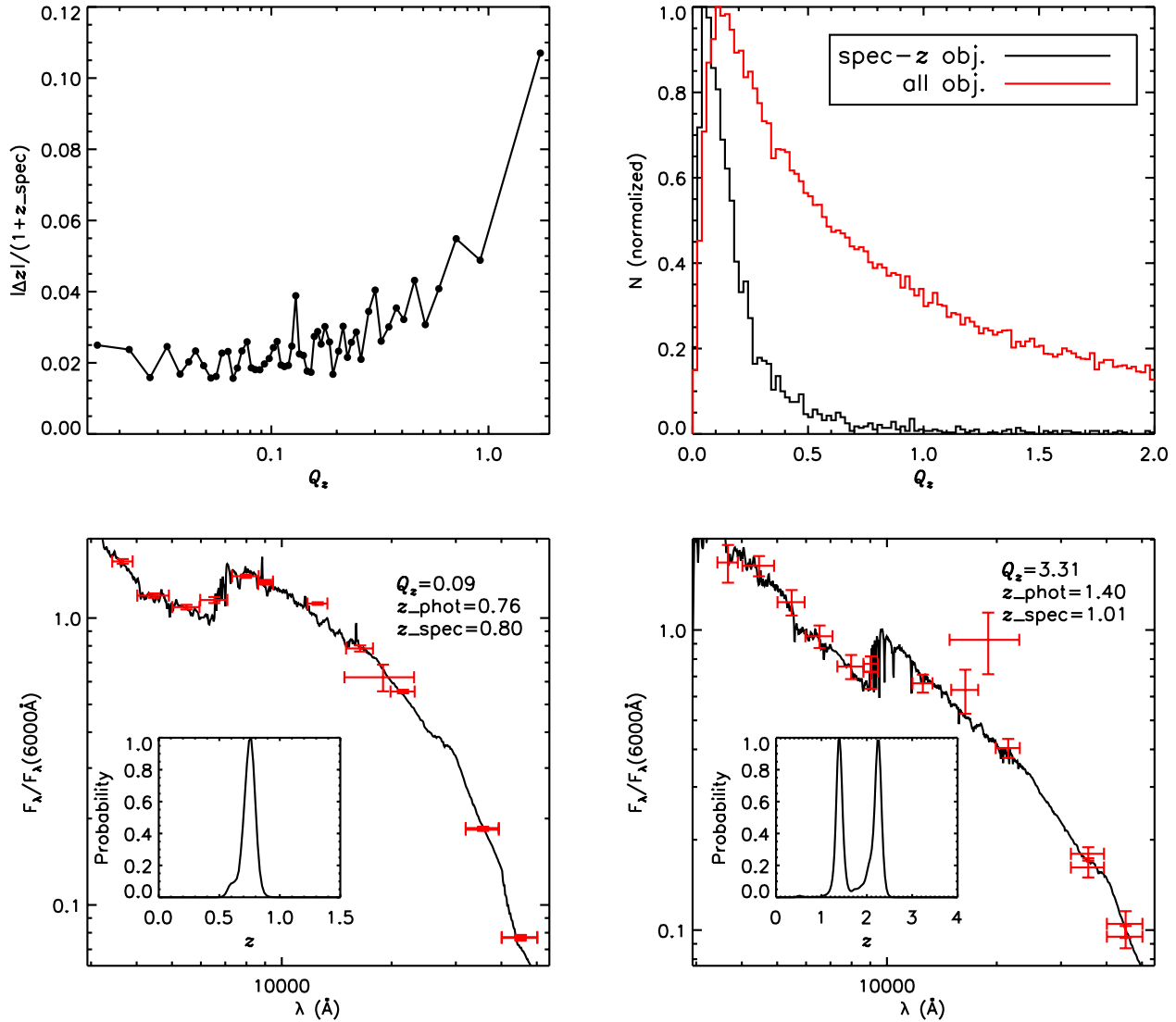}
\caption{(Top-left) Plot of $\rm |\Delta \z|/(1+\zs)~versus~\it Q_z$, where each 50 sources are binned into one data point. (Top-right) Histograms of $Q_z$ for the \zs~sources (black curve) and all sources (red curve). (Bottom-left) A typical best-fit SED template for a low-$Q_z$ source. (Bottom-right) A typical best-fit SED template for a high-$Q_z$ source. The insets in the bottom panels show the \zp\ probability distributions with the peaks being normalized to unity. \\(A color version of this figure is available in the online journal.)}\label{fig:qz}
\end{figure*}

Here we adopt the following criterion for source selection. First we calculate the redshift quality parameter $Q_z$ (see Equation~8 of B08) for each source in multiple catalogs as 
\begin{equation}
Q_z=\frac{\chi^2}{N_{\rm filt}-3}
\frac{\z _{\rm up}^{99}-\z _{\rm lo}^{99}}{P_{\delta z =0.2}},
\end{equation}
where $\chi^2$ is obtained from template fitting; $N_{\rm filt}-3$ is the degrees of freedom; $\z _{\rm up}^{99}-\z _{\rm lo}^{99}$ is the 99\% confidence level interval that represents the \zp~scatter (Mobasher et al. 2007); and $P_{\delta \z}$ is the fraction of the total integrated probability that lies within $\pm (1+\zp)\delta z$ of the \zp~estimate, designed to identify sources that have broad and/or multi-modal probability distributions (Ben\'itez 2000). Then if a source appears in more than one catalog (i.e., multiple detections), we regard the detection with the lowest $Q_z$ value as the most reliable one and adopt its corresponding photometry and \zp. The relation between $Q_z$ and  $|\Delta z|/(1+\zs)$ (where $\Delta z=\zp -\zs$) is plotted in the top-left panel of Figure~\ref{fig:qz}. In Figure~\ref{fig:qz}, we also plot $Q_z$ histograms for spectroscopic and all sources, respectively, in the top-right panel (as expected, the spectroscopic sources typically have smaller $Q_z$ values indicating better \zp~quality than average), and present two typical SED fitting results that correspond to two $Q_z$ values in the two bottom panels. The first SED is fitted well by the template with $Q_z=0.09$, and its \zp~probability distribution shows a single peak. In contrast, the second SED is fitted relatively poorly by the template with $Q_z=3.31$, and its \zp~probability distribution shows double peaks.

\begin{figure}
\center
\includegraphics[width=\linewidth]{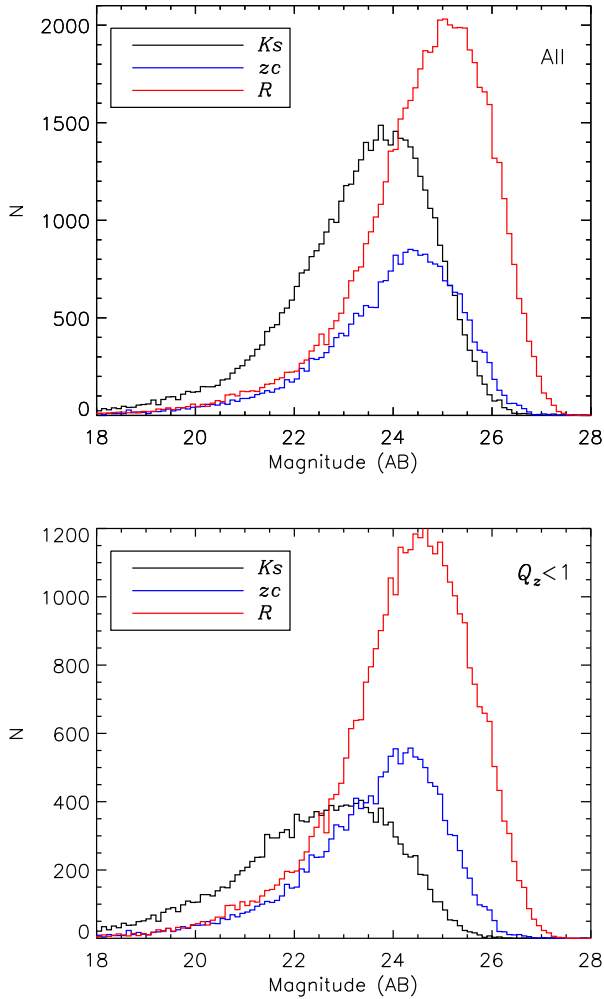}
\caption{Histograms of magnitudes in the detection bands for all sources (top panel) and those with \hbox{$Q_z < 1$} (bottom panel). Sources detected in the \bandk, \bandzc, and \bandr~bands are plotted as black, blue, and red curves, respectively.\\(A color version of this figure is available in the online journal.)}\label{fig:mag}
\end{figure}

Following the above source-selection criterion, we obtain a total of 131,678 distinct sources, among which 46447, 25478, and 59753 sources are selected from the \bandk, \bandzc, and \bandr\ catalogs, respectively. Typically, a value of \hbox{$Q_z \lsim 1$} indicates reliable \zp~quality (i.e., $|\Delta z|/(1+\zs)\lsim 0.05$). There are 67,415 sources with \hbox{$Q_z < 1$} in our merged catalog, among which 15,322, 16,224, and 35,869 sources are detected in the \bandk, \bandzc, and \bandr~bands, respectively.
Figure~\ref{fig:mag} shows histograms of magnitudes in the detection bands for all sources and those with \hbox{$Q_z < 1$}, respectively. Typically, sources with lower $Q_z$ values tend to be brighter, e.g., the median magnitudes are \bandk=22.6, \bandzc=23.8, and \bandr=24.4 mag for sources with \hbox{$Q_z < 1$}, and \bandk=23.5, \bandzc=24.1, and \bandr=24.8 mag for all sources, respectively.

\subsection{\zp~Quality}\label{sec:zp qua}

\begin{figure*}
\includegraphics[width=\linewidth]{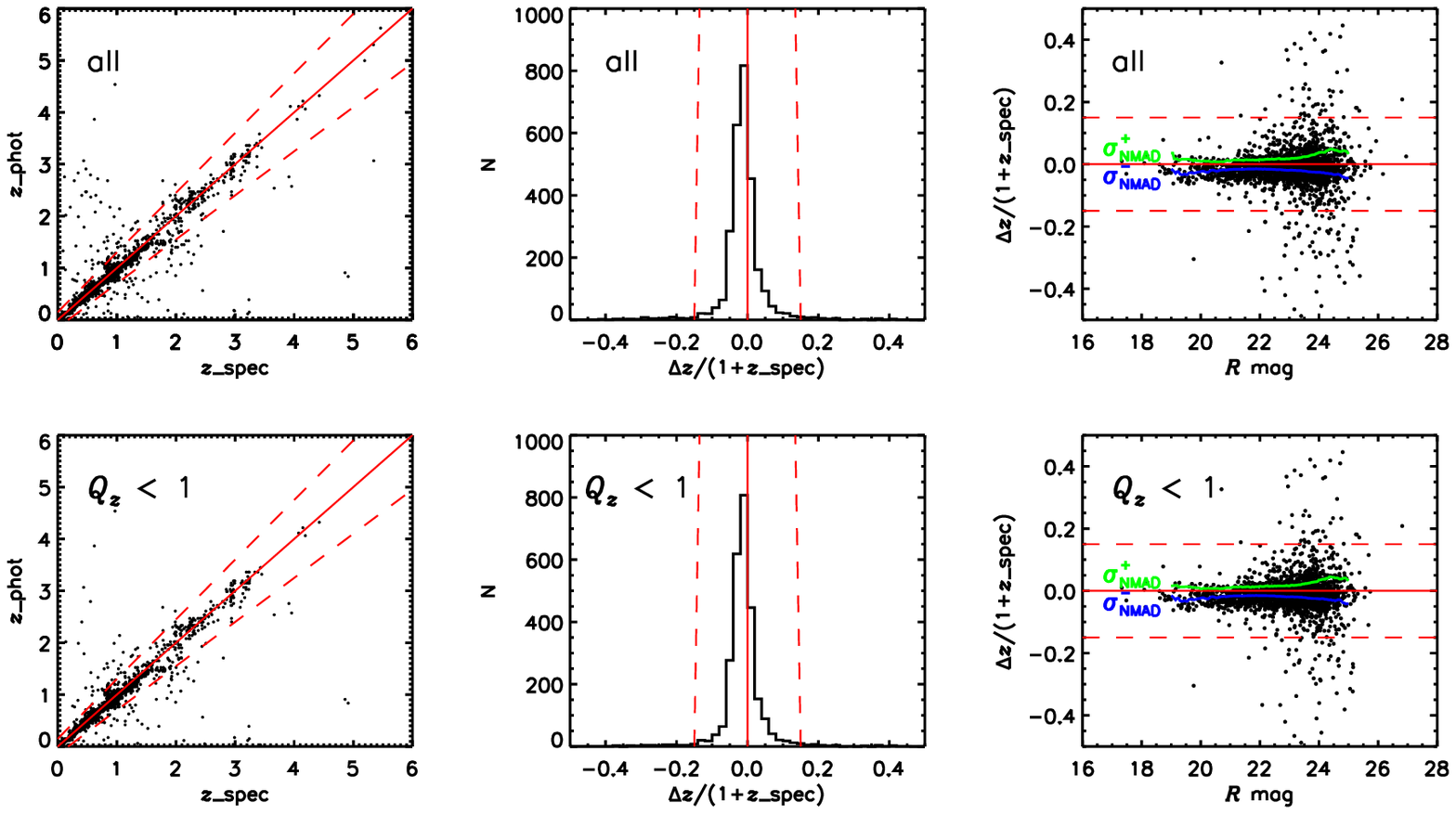}
\caption{Plots of \zp~versus~\zs~(Left), histograms of $\Delta \z/(1+\zs)$ (Middle), and plots of $\Delta \z/(1+\zs)$ versus \bandr-band magnitude (Right) for all non-X-ray \zs~sources (2845 in total; the top row) and all non-X-ray \zs~sources with $Q_z<1$ (2744 in total; the bottom row).
Red solid lines indicate $\rm\ \Delta \z/(1+\zs) =0$ and red dashed lines indicate $\rm\ \Delta \z/(1+\zs) = \pm0.15$. The $\rm \sigma_{NMAD}^+$ and $\rm \sigma_{NMAD}^-$ running curves are computed according to Equation~\ref{equ:sigma} for sources with $\rm\ \Delta \z/(1+\zs) > 0$ and $\rm\ \Delta \z/(1+\zs) < 0$ (in bins of $\rm \Delta \bandr=1~mag$) and shown as green and blue curves, respectively.\\(A color version of this figure is available in the online journal.)}
\label{fig:res}
\end{figure*}

We find 2845 matches between our sources and the \zs\ catalogs for non-X-ray objects (see Table~\ref{tab:zspec}; also see Section~\ref{sec:zp cor} for our matching approach).
Then we use the \zs\ data to evaluate our \zp\ quality (for evaluation of \xray\ objects, see Section~\ref{sec:agn}). We obtain $\rm \sigma_{NMAD}=0.029$ for those 2845 sources. The median value of $\rm \Delta \z/(1+\zs)$ is $-0.013$ and there are 156 ($156/2845=5.5\%$) outliers. 
More specifically, we find \sigm$=0.024$ with 2.7\% outliers for sources brighter than $R=23$~mag, \sigm$=$0.035 with 7.4\% outliers for sources fainter than $R=23$~mag, \sigm$=$0.026 with 3.9\% outliers for sources having $z<1$, and \sigm$=$0.034 with 9.0\% outliers for sources having $z>1$. Figure~\ref{fig:res} demonstrates our \zp~quality. The top three panels are, from left to right, \zp~vs.\ \zs, histogram of $\rm \Delta \z/(1+\zs)$, and $\rm \Delta \z/(1+\zs)$ vs.\ \bandr-band magnitude for the 2845 sources, respectively. The bottom three panels are the same as the top panels but are limited to the sources with $Q_z<1$. From comparison between the top and bottom panels, we find the criterion of $Q_z<1$ filters out many outliers. The \zp~quality is also related to the number of available bands, e.g., $\rm |\Delta \z|/(1+\zs)\sim 0.041$ when 12 bands (upper limits not counted) are available while $\rm |\Delta \z|/(1+\zs)\sim 0.027$ when all 15 bands are available.

We compare our work with Rafferty et al. (2011), which is the most relevant work where they estimated \zp~for sources in the central \hdfn\ area based on broadband photometry from C04 and some other broadband photometry, using the ZEBRA template-fitting code (Feldmann et al.\ 2006).
The major differences of our work from Rafferty et al. (2011) are
(1) we include additional high-quality data that have become available only recently (e.g., the \bandk~band from W10 as well as the IRAC data from the Spitzer Heritage Archive and Ashby et al. 2013);
(2) we derive uniform PSF-matched photometry rather than compiling a number of magnitudes of various origins and different derivation methodologies;
(3) our catalog consists of 131,678 sources, many more than the 48,858 sources in Rafferty et al. (2011), due to our larger solid-angle coverage (i.e., the entire \hdfn\ field), deeper data, different catalog-construction approach, and inclusion of lower-significance sources;
and (4) our \zp\ estimation does not involve apparent training procedures. 
Rafferty et al. (2011) reached \sigm$=0.025$ and an outlier fraction of 5.0\% (outliers defined as having $\rm |\Delta \z|/(1+\zs) >0.20$ therein) with training procedures applied. However, their blind-test results showed that their real \zp~quality should be worse by a factor of a few, indicating that our \zp\ quality has an overall improvement over their work.

\section{The Final Catalog}\label{sec:cat}

\subsection{Absolute Photometry}\label{sec:abs phot}
Our approach of PSF-matched photometry extraction is designed to obtain accurate colors rather than absolute fluxes, and it underestimates the fluxes because the aperture diameter is fixed at 1.5 times the PSF size ($\sim70\%$ of light encircled for point-like sources) for all sources in each band. To convert the aperture photometry to the absolute one, we make use of FLUX\_AUTO in SExtractor's output of the detection band, i.e., the \bandk, \bandzc, or \bandr~band. The algorithm of FLUX\_AUTO adopts a flexible aperture size for each source (Kron et al.\ 1980), and FLUX\_AUTO has been widely used to obtain absolute photometry (e.g., Cardamone et al.\ 2010; W10). If a source has detection in the X (representing \bandk, \bandzc, or \bandr) band, then we obtain absolute flux of Y (representing any of the 17 bands), $f_{\rm Y, final}$, as  
\begin{equation}
f_{\rm{Y}, final}=c\times f_{\rm{Y,aper}}\times \frac{f_{\rm{X,AUTO}}}{f_{\rm{X,aper}}},
\end{equation}   
where $c$ is the correction factor to convert FLUX\_AUTO to absolute flux ($c=1.06$ in our case, according to Page 39 of the SExtractor manual for version 2.13). The above procedure conserves colors, which indicates that the same \zp\ results would be obtained with either the relative or absolute photometry.

We compare our \bandk-band absolute photometry (without applying zero-point offsets) with that of W10 for the common sources that have $\ge 5\sigma$ significance, and find a median offset of $\sim 0.10$ mag and a scatter of 0.15\ mag. No straightforward comparison can be made between our absolute photometry and that of C04, because C04 adopted a different approach to obtain absolute photometry, i.e., using isophotal fluxes of SExtractor rather than auto fluxes that we adopted. 

\subsection{Absolute Astrometry}\label{sec:abs astrom}
In the astrometry correction procedure (see Section~\ref{sec:astrom}), we align all images to the astrometric frame of the C04 images that were well aligned with each other thus being optimal for obtaining accurate colors. However, this astrometry might have subtle systematic errors. To account for this, we match our sources with those detected by the VLA 1.4~GHz observations (Morrison et al.\ 2010) using a 1.0\arcsec~matching radius. Then we use {\sc geomap} and {\sc geoxytran} (IRAF tasks) to correct our astrometry so that it is more consistent with that of the VLA data. As in Section~\ref{sec:astrom}, we also adopt a 4th-order polynomial correction that is applied to the entire catalog. The median values of the coordinate offsets applied are 0.28\arcsec\ (RA) and $-$0.17\arcsec (DEC).

\subsection{Star/Galaxy Classification}\label{sec:star}
We classify a source by fitting its photometry with the set of 235 stellar templates introduced in Section~\ref{sec:zp cor} at \z =0 and another set of 259 galaxy templates at \z~=\zp.
The galaxy templates are the P\'EGASE2.0 templates (259 in total; Grazian et al.\ 2006) taken from the EAzY package. For the purpose of star/galaxy separation, the fitting is done via the single template mode (STM) rather than the linear combination mode (LCM) of EAzY, because the former yields slightly more consistent results with the \textit{BzK} method (Daddi et al.\ 2004) that separates effectively stars from galaxies. In this mode (STM), we use a 5\% error in place of the template error function, since the template error function is designed for the default template set in the LCM. We discard all IRAC data for the purpose of star/galaxy separation because the stellar templates of Pickles (1998) do not cover wavelengths longer than 2.5~$\mu$m. We classify a non-spectroscopic source as a star only if it satisfies the following two criteria: 
(1) $\chi_{\rm star}^{2}<\chi_{\rm gal}^{2}$; and
(2) if the source has S/N$>5$, we then require additionally that its major axis to minor axis ratio (from SExtractor) be in the range of 1.0--1.5;  
where $\chi_{\rm star}^{2}$ is the fitted $\chi^{2}$ using the 235 stellar templates at \z =0 and $\chi_{\rm gal}^{2}$ is the fitted $\chi^{2}$ using the 259 galaxy templates at \z~=\zp.

\begin{figure}
\includegraphics[width=\linewidth]{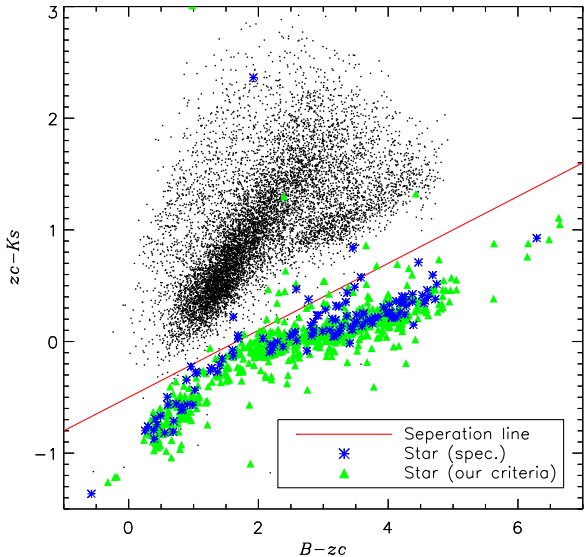}
\caption{\textit{BzK} map for star/galaxy classification. Only sources with reliable $B$, $zc$, and $K_s$ photometry (i.e., $\rm err\_mag_\bandb+err\_mag_\bandzc < 0.5$ and $\rm err\_mag_\bandzc+err\_mag_\bandk < 0.15$) are plotted as black dots. Spectroscopically-identified stars are marked as blue asterisks; candidate stars selected by our template-fitting and morphological criteria are marked as green filled triangles.\\(A color version of this figure is available in the online journal.)}
\label{fig:bzk}
\end{figure}

Finally, our method identifies a total of 4959 star candidates (with the 229 spectroscopic stars being counted). Among them, 25 are best-fitted by a white dwarf template. To verify the accuracy of our method, we also make a \textit{BzK} diagram. In Figure~\ref{fig:bzk}, we only plot sources that have reasonably accurate photometry in the relevant three bands (i.e., $\rm err\_mag_\bandb+err\_mag_\bandzc < 0.5$ and $\rm err\_mag_\bandzc+err\_mag_\bandk < 0.15$). We find that our template fitting method is consistent with the \textit{BzK} star/galaxy classification scheme. As expected, if the sources with larger photometric errors are also plotted in Figure~\ref{fig:bzk}, then the
star/galaxy separation is not as good, but it is still reasonable.
For spectroscopic sources, we misclassified 13 galaxies as stars (out of a total of 3126; 13/3126=0.4\%) and 24 stars as galaxies (out of a total of 229; 24/229=10.5\%) according to the above criteria. Of those 24 misclassified stars, about one half are very bright thus suffering from issues of bad pixels and/or saturation, and the other half or so are blended with nearby sources to some degree.

\subsection{AGNs}\label{sec:agn}
We match our sources (using the optical and near-infrared positions as well as the \bandk-band magnitude distribution) with the 2~Ms CDF-N (A03; see Figure~\ref{fig:field} for coverage) main \xray\ source catalog (using \xray\ positions) utilizing the likelihood-ratio matching technique presented in Luo et al. (2010). There are 462 non-stellar X-ray sources matched (13 additional \xray\ stars matched), with 281 having \zs\ (see Table~\ref{tab:zspec}). The false matching rate is estimated to be $\approx 2\%$. The \zp~quality of those \xray\ sources is as follows: \sigm$=0.037$, an outlier fraction of 16.7\% (i.e., 47 outliers), and $\rm median(\Delta \z / (1+\zs))=-0.014$; such \zp\ quality is worse than that of non-X-ray sources (see Section~\ref{sec:zp qua}). 

There are two main reasons for the worse \zp~quality of \xray\ sources. First, the vast majority ($\gsim$75\%; based on studies of deep \xray\ surveys; e.g., A03; Luo et al. 2008; Xue et al. 2011) of these \xray\ sources are active galactic nuclei (AGNs). Typically, many AGNs have different SEDs to normal galaxies and therefore it may not be correct to estimate \zp~for AGNs with normal galaxy templates, especially in the case of QSOs whose SEDs are dominated by the central engine. Second, fluxes of AGNs might vary non-periodically on timescales from minutes to decades (e.g., Salvato et al.\ 2009). Salvato et al.\ (2009) took AGN variability into account when deriving \zp\ based on multi-epoch observations (i.e., one filter has observations in different epochs). However, all our bands were not observed in the same epoch (see Table~\ref{tab:img}) and we do not have multi-epoch data for a specific band. Therefore our photometry might differ from a snapshot SED, thus being likely to be subject to uncertainties due to AGN variability.
 
\begin{figure*}
\includegraphics[width=\linewidth]{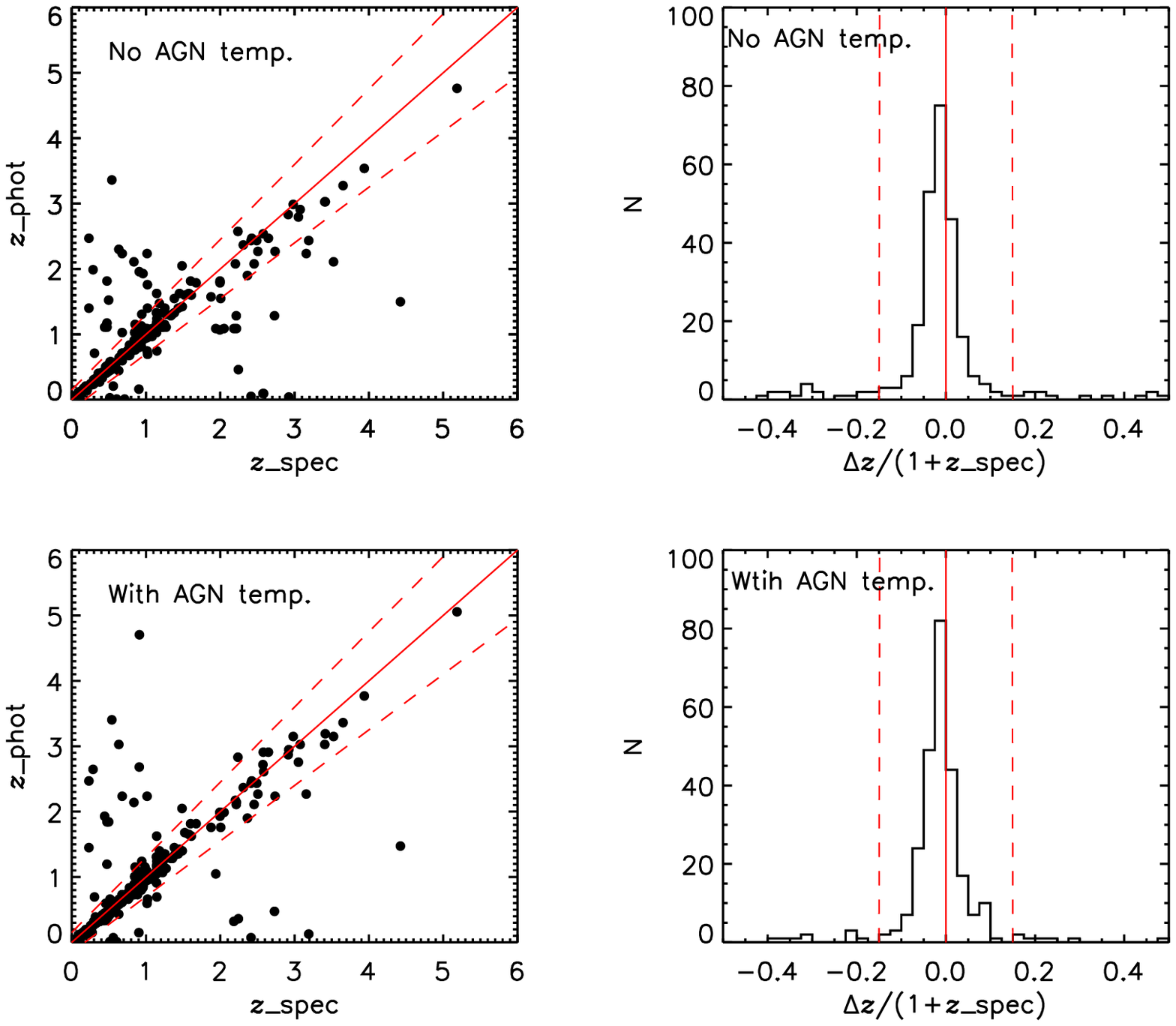}
\caption{The top two panels are plots of \zp~versus~\zs~and the histogram of $\rm\ \Delta \z/(1+\zs)$ for \xray~sources (281 in total) before adding AGN templates when estimating \zp. The bottom two panels are the same as the top two panels but after adding AGN templates. Red solid lines indicate $\rm\ \Delta \z/(1+\zs) =0$ and red dashed lines indicate $\rm\ \Delta \z/(1+\zs) = \pm0.15$.\\(A color version of this figure is available in the online journal.)}
\label{fig:agn}
\end{figure*}

To improve the \zp~quality of \xray\ sources, we add three additional QSO templates to the previous template set when estimating \zp\ for \xray\ sources. The first two are the BQSO and TQSO templates from the SWIRE library (Polletta et al.\ 2007), both of which are type 1 QSO but with different IR/optical flux ratios; the third one is the optical-to-near-infrared composite QSO template from Glikman et al.\ (2006), and is built from spectra of QSOs at different redshifts. As in Section~\ref{sec:star}, we use a 5\% error instead of the template error function, because the template error function is designed for normal galaxies rather than AGNs (B08). Other EAzY parameters are the same as those in Section \ref{sec:oth}. Notably, the LCM algorithm naturally mixes the QSO and stellar light. The resulting \zp~quality is improved appreciably: \sigm$=0.035$, an outlier fraction of 12.5\% (i.e., 35 outliers), and $\rm median(\Delta \z / (1+\zs))=-0.014$. Figure~\ref{fig:agn} demonstrates this improvement in \zp~quality by showing plots of \zp~quality before and after the introduction of the above three QSO templates. Overall, our \zp~quality of X-ray sources appears comparable to those presented in other similar works, e.g., Rafferty et al. (2011). 

It can also be seen in Figure~\ref{fig:agn} that a small fraction of \xray\ sources that were fit well (and thus have good \zp) without adding the QSO templates are no longer fit well when adding the QSO templates, and vice versa. This is due to degeneracy.
To examine the effect of degeneracy introduced by adding these three QSO templates, we compare \zp\ results obtained for the 462 \xray\ sources, using only the eight galaxy templates shown in Figure~\ref{fig:tem} and using both the eight galaxy templates and the three QSO templates, respectively. We find a nominal \sigm~$=0.015$ and a nominal outlier fraction of 6.5\%, indicating that the effect of this degeneracy is insignificant.

\subsection{Advice on Using Our Catalog}\label{sec:guide}
The users of our catalog should be careful given that we do not apply a significance cut to the sources that are included in our catalog, which might lead to some false detections. Instead, we choose to provide both the significance level (i.e., S/N) and the redshift quality parameter $Q_z$ in our final catalog (see Section~\ref{sec:catdet}), and recommend the users to apply an appropriate S/N and/or $Q_z$ cut according to their specific scientific interests. We note that, as expected, there is a general anti-correlated trend between S/N and $Q_z$ in spite of significant scatter. 

Additionally, the users of our catalog should exercise extra caution in making use of the sources lying in the extended \hdfn~field (i.e., the regions outside the yellow rectangle in Figure~\ref{fig:field}), given that, for the sake of completeness, we provide all the sources in the entire \hdfn~field in our catalog, rather than only those lying within the C04 central \hdfn~field. The sources in the extended field might potentially suffer from issues such as additional astrometry distortions and large background noise (see Figure~\ref{fig:field}). The latter could cause large photometric errors and additional false detections (C04). Unfortunately, it is rather difficult to assess \zp\ quality in the extended field where there are not any \zs\ data. We thus provide a flag indicating whether a source is located in the central or extended \hdfn~field in our final catalog (see Section~\ref{sec:catdet}) so that the users could proceed with known caveats.

\subsection{Catalog Details}\label{sec:catdet}
The \hdfn~photometric catalog is presented in Table~\ref{tab:cat}. The details of the 53 columns are given below. 

 1.\ Column 1 gives the source sequence number (i.e., ID, ranging from 1 to 131,678). We list sources in order of increasing right ascension. 

 2.\ Columns 2 and 3 give the J2000.0 right ascension and declination, respectively. They are consistent with the VLA radio astrometry (see Section~\ref{sec:abs astrom}).  

 3.\ Column 4 gives the \zp~value, which corresponds to the highest peak in the corresponding \zp~probability distribution output by EAzY.
 
 4.\ Column 5 gives the alternative \zp~value denoted as \za\ if available. In the \zp~probability distribution, if a source has other peaks in addition to the highest one (corresponding to \zp) and satisfies the following two conditions, i.e.,
  \begin{eqnarray} \label{equ:za}
     \frac{|\ze - \zp|}{1 + \zp} > 0.15, {\rm and}\\
     p(\ze) > 0.5 \times p(\zp),
     \end{eqnarray}
   where $p(\ze)$ and $p(\zp)$ are the values in the probability distribution quoted at redshifts of \ze~and \zp, respectively; we then define \za~as the redshift that corresponds to the highest peak among those peaks. 34\% of all sources in our catalog have \za~values, while for sources with \hbox{$Q_z < 1$} the fraction is 13\%. We set $\za=0$ for stars and $\za=-1$ for sources whose \za\ values are not available.

  5.\ Columns 6--11 give the 68\% ($1\sigma$), 95\% ($2\sigma$), and 99.7\% ($3\sigma$) lower and upper limits on \zp, respectively, which are calculated as
  \begin{eqnarray}\label{equ:limit}
    \alpha/2 &=& \int_0^{z_\mathrm{low}} \! p(z) \, \mathrm{d}z,\\
     \alpha/2 &=& \int_\mathrm{\z_{up}}^8 \! p(z) \, \mathrm{d}z,\nonumber 
  \end{eqnarray}
  where $\alpha=$0.317, 0.046, and 0.003, respectively, and $p(z)$ is the probability distribution of redshift. These limit values are output by EAzY. Note that our \zp~corresponds to the peak value of $p(z)$, thus in some cases $z_{\rm low}$ might be greater than \zp, or $z_{\rm up}$ might be lower than \zp.\footnote{To give an extreme example, if a source has $\zp=0.01$ (i.e., the \zp\ probability distribution peaks at the minimum \zp~grid value of 0.01), then its $z_{\rm low}$ is very likely to be greater than \zp, because $z_{\rm low}$ has to be large enough to make the integral reach $\alpha/2$ in the left-hand side of Eq.~\ref{equ:limit}.} A total of $\sim 7\%$ of sources in our catalog have $\rm \z_{low}>\zp$ or $\rm \z_{up}<\zp$. For those sources, the abnormal $\rm \z_{low}$ or $\rm \z_{up}$ value should not be used. 

 6.\ Column 12 gives the source type: values of $-2$, $-1$, 0, and 1 indicate white dwarfs (25 sources), other stars (4934 sources; see Section~\ref{sec:star}), non-\xray~sources (126,257 sources), and \xray~sources (462 sources; see Section~\ref{sec:agn}), respectively. If a source is classified as a star, then we do not count it as an \xray~nor non-\xray~source, set its \zp\ and corresponding confidence ranges as 0, and set $Q_z$ as $-1$. We refer whoever interested in \xray\ stars (a total of 14 in our catalog) to both this column and Column 18.
 
 7.\ Column 13 gives the redshift quality parameter $Q_z$ (see Section~\ref{sec:qz}). Generally, lower $Q_z$ values indicate better \zp~quality. We suggest a criterion of \hbox{$Q_z < 1$} for reliable \zp. There are 67,415 sources with \hbox{$Q_z < 1$} in our catalog.
   
 8.\ Column 14 gives the detection significance or signal-to-noise ratio (S/N), defined as 
      \begin{equation}\label{equ:sn}
      \rm S/N = \frac{flux} {background\ noise},
      \end{equation}
      where flux and background noise (see Section~\ref{sec:phot err}) are for the detection band (see Section~\ref{sec:qz}). The users of our catalog are recommended to apply an appropriate cut based on $Q_z$ and/or S/N to filter out sources of poor photometry and thus \zp\ quality, in order to fulfill effectively their specific science goals.

 9.\ Column 15 gives the detection band: letters of K, Z, and R indicate \bandk-, \bandzc-, and \bandr-band detections, respectively. If a source is detected in multiple bands, the first letter indicates the adopted detection catalog according to the lowest $Q_z$ criterion (see Section~\ref{sec:qz}). After applying this criterion, the numbers of sources selected from the \bandk, \bandzc, and \bandr\ catalog are 46447, 25478, and 59753, respectively. 
 
 10.\ Columns 16 and 17 give the \zs~value and its reference index, respectively, if available. We only include secure \zs\ in this work. The reference indexes are listed in Table~\ref{tab:zspec}. Situations where a source has more than one reference are dealt with in Section~\ref{sec:zs cat}. If \zs~for a source is not available, then both values are set to $-1$. There are a total of 3355 sources having \zs, including 2845 non-\xray~sources, 281 \xray~sources, and 229 stars.
 
 11.\ Column 18 gives the source index in the A03 2~Ms \cdfn~main catalog if it has a match therein (see Section~\ref{sec:agn}). For the sources not matched to A03, the value of this column is set to $-1$. The number of sources matched to A03 is 475, including 462 non-stellar \xray\ sources and 13 \xray\ stars.

 12.\ Column 19 gives a flag indicating whether the source is in the C04 central \hdfn\ region: 0 stands for being outside of the central region (i.e., in the extended region; 53,636 sources) and 1 stands for being in the central region (78,042 sources). The properties of the sources in the central region are of better overall quality than those outside (see Section~\ref{sec:complete}).

 13.\ Columns 20--53 give the photometry and corresponding errors in magnitudes.
This is the corrected photometry derived in Section~\ref{sec:zp cor} that has further been corrected to the absolute photometry based on FLUX\_AUTO algorithm in SExtractor (see Section~\ref{sec:abs phot}). We do not apply the zero-point offsets derived in Section~\ref{sec:zp} in the final catalog. The order is \bandu~band, error of \bandu~band, \bandb~band, error of \bandb~band, and so forth (the band order is the same as that in Table~\ref{tab:img}). If the photometry of a source in one band is not available (due to, e.g., being outside of the field of view or saturated), then we set corresponding columns to values of $-99$. If the flux is less than its error (both in units of $\mu$Jy), we apply an upper limit, i.e., set the flux to the value of the error (see Section~\ref{sec:phot err}). For completeness, the \sthr- and \sfor-band photometry, which is not used in \zp~calculation (see Section~\ref{sec:zp}), is also presented.

\begin{table*} 
\begin{center}
\caption{H-HDF-N photometric catalog}\label{tab:cat}
\begin{tabular}{c c c c c c c c c c c c c c c c c c c}\hline\hline
ID & $\rm \alpha_{J2000.0}$  & $\rm \delta_{J2000.0}$ & \zp & \za & L68 & U68 & L95 & U95 & L99 & U99 & Type & $Q_z$ & S/N & Detect & \zs \\
(1) & (2) & (3) & (4) & (5) & (6) & (7) & (8) & (9) & (10) & (11) & (12) & (13) & (14) & (15) & (16) \\ \hline
1 & 188.52433 & 62.348192 & 0.000 & 0.000 & 0.000 & 0.000 & 0.000 & 0.000 & 0.000 & 0.000 & $-1$    & $-1.000$    & 16.3   & Z   & $-$1.000  \\
2 & 188.52454 & 62.299627 & 0.184 & 1.152    & 0.411 & 3.259 & 0.080 & 4.796 & 0.013 & 6.185 & 0    & 33.2     & 3.21  & Z & $-$1.000 \\   
3 & 188.52457 & 62.306804 & 0.000 & 0.000 & 0.000 & 0.000 & 0.000 & 0.000 & 0.000 & 0.000 & $-1$    & $-1.000$     & 11.6   & Z & $-$1.000 \\
4 & 188.52474 & 62.379926 & 0.138 & 2.682    & 0.245 & 2.562 & 0.044 & 2.792 & 0.011 & 2.910 & 0    & 6.19     & 2.54   & RZ & $-$1.000 \\
5 & 188.52486 & 62.334587 & 0.694 & $-$1.000 & 0.383 & 1.012 & 0.060 & 1.931 & 0.011 & 2.650 & 0    & 0.163    & 4.91   & Z & $-$1.000 \\ \hline  
\end{tabular}
\end{center}
{\sc Note.} --- 
The full table contains 53 columns of information for the 131,678 sources (see Section~\ref{sec:catdet} for the descriptions of the columns.)\\
(This table is available in its entirety in a machine-readable form in the online journal. A portion is shown here for guidance regarding its form and content.)
\end{table*}

\section{summary}\label{sec:sum}
Following the procedures outlined in Figure~\ref{fig:flow}, we have derived \zp~for 131,678 sources in the entire \hdfn~region (including both the central and extended areas) based on 15 broadband images, i.e., the \bandu, \bandb,  \bandv,  \bandr,  \bandi,  \bandzc,  \bandzo, \bandj, \bandh, \bandk, \bandhk, \aone, \sone, \atwo, and \stwo\ bands. PSF-matched photometry is extracted in order to obtain accurate colors that are the key to achieving high-quality \zp, given that the PSFs of our imaging data differ significantly. We compute the zero point of each band to eliminate systematic offsets and then estimate \zp\ with EAzY (however, we do not apply the zero-point offsets to the photometry presented in our final catalog). Two additional galaxy templates are added to the EAzY default template set in order to obtain \zp~with higher accuracy. We classify the sources in our catalog as stars or galaxies based on SED fitting and complementary morphological parametrization. Furthermore, we match our sources with the A03 2~Ms \cdfn\ main-catalog \xray\ sources using a likelihood-ratio matching technique, resulting in 462 non-stellar \xray\ sources. To evaluate our \zp~quality, we compare our \zp~with \zs~when available and find \sigm$=$0.029 with an outlier fraction of 5.5\% for the 2845 non-\xray\ spectroscopic galaxies. More specifically, we find \sigm$=0.024$ with 2.7\% outliers for sources brighter than $R=23$~mag, \sigm$=$0.035 with 7.4\% outliers for sources fainter than $R=23$~mag, \sigm$=$0.026 with 3.9\% outliers for sources having $z<1$, and \sigm$=$0.034 with 9.0\% outliers for sources having $z>1$. This \zp~quality is comparable to those presented in previous similar works. The above \zp~procedure yields a relatively poor \zp~quality for \xray~sources (281 \xray~sources have \zs), with \sigm$=0.037$ and an outlier fraction of 16.7\%. To improve this situation, we add three additional AGN templates, and obtain an improved \zp~quality for \xray\ sources, with \sigm$=0.035$ and an outlier fraction of 12.5\%. We make our catalog publicly available and provide guidance on how to make use of it.

\acknowledgments

We thank the referee for helpful feedback that improved this work.
We thank Matthew Ashby for advice on the use of SEDS data, Ryan Keenan for providing the \bandj- and \bandh-band images, Wei-Hao Wang for providing information on the \bandk-band image, and Masami Ouchi for providing the \bandzo-band image. We are grateful to Gustavo Bruzual for providing the newest version of GALAXEV, to Russ Laher for his novel APT software, to Roberto Assef for the Low Resolution Templates code, and to Gabriel Brammer for the EAzY code.    
We also thank Tinggui Wang for helpful discussions.
G.Y. and Y.Q.X. acknowledge the financial support of the Thousand Young Talents (QingNianQianRen) program (KJ2030220004), the 973 Program (2015CB857004), the USTC startup funding (ZC9850290195), the National Natural Science Foundation of China (NSFC-11473026, 11421303, 11243008), the Strategic Priority Research Program ``The Emergence of Cosmological Structures'' of the Chinese Academy of Sciences (XDB09000000), and the Fund for Fostering Talents in Basic Science of the National Natural Science Foundation of China (NSFC-J1310021).
W.N.B. and B.L. acknowledge support from the CXC grant AR3-14015X and the NASA ADP grant NNX10AC99G. 
D.M.A acknowledges support from the UK Science and Technology Facilities Council (STFC, grant ST/I001573/I) and the Leverhulme Trust.
F.E.B. acknowledges support from Basal-CATA PFB-06/2007, CONICYT-Chile (FONDECYT 1141218, Gemini-CONICYT 32120003, ``EMBIGGEN'' Anillo ACT1101), and Project IC120009 ``Millennium Institute of Astrophysics (MAS)'' funded by the Iniciativa Cient\'{\i}fica Milenio del Ministerio de Econom\'{\i}a, Fomento y Turismo.


\end{document}